\documentclass[fleqn,usenatbib]{mnras}

\usepackage[T1]{fontenc}

\DeclareRobustCommand{\VAN}[3]{#2}
\let\VANthebibliography\thebibliography
\def\thebibliography{\DeclareRobustCommand{\VAN}[3]{##3}\VANthebibliography}

\usepackage{graphicx}	% Including figure files
\usepackage{amsmath}	% Advanced maths commands
\usepackage{amssymb}	% Extra maths symbols
\usepackage{siunitx}    % Package included by me for scientific notation
\usepackage{orcidlink}
\usepackage{soul}
\usepackage{newtxtext,newtxmath}
\title[The EKM as the cause of transits of KIC 8462852]{The Eccentric Kozai-Lidov Mechanism as the Cause of Exocomet Transits of KIC 8462852}

\author[S. D. Young and M. C. Wyatt]{
    Steven. D. Young \orcidlink{0000-0002-4008-323X} $^{1}$ \thanks{E-mail: sdy21@cam.ac.uk} and 
Mark. C. Wyatt \orcidlink{0000-0001-9064-5598} $^{1}$
\\
$^{1}$Institute of Astronomy, University of Cambridge, Madingley Road, Cambridge, CB3 0HA, UK\\
}

\date{Accepted XXX. Received YYY; in original form ZZZ}

\pubyear{2021}

\begin{document}
\label{firstpage}
\pagerange{\pageref{firstpage}--\pageref{lastpage}}
\maketitle

\begin{abstract}
KIC 8462852 is a star in the Kepler field that exhibits almost unique behaviour. The deep, irregular and aperiodic dips in its light curve have been interpreted as the breakup of a large exocomet on a highly eccentric orbit whose post-disruption material obscures the star. It is hypothesised that a nearby M-dwarf, recently confirmed to be bound to the system, could be exciting planetesimals in a source belt to high eccentricities if its orbit is highly misaligned with the belt: an effect known as the `Eccentric Kozai-Lidov Mechanism'. To quantify how often this effect is expected to occur, this paper presents a Monte Carlo model of wide binary stars with embedded, misaligned planetesimal belts. These belts collisionally erode over time until they are excited to high eccentricities on secular timescales by a companion star if its orbit is sufficiently misaligned. The large planetesimals then produce an observable dimming signature in the light curve for a set period of time which may or may not overlap with similar events. The model finds that, for dimming events that persist for 100 yr, the most likely companion stars are located at $10^2 - 10^4$ au, the most likely belts are at $10^2-10^3$ au and the system age is most likely to be $10^2 - 10^3$ Myr. However, the probability of observing one or more stars exhibiting this phenomenon in the Kepler field is $\num{1.3e-3}$, such that it is unlikely this mechanism is driving the observations of KIC 8462852.
\end{abstract}


%%%%%%%%%%%%%%%%%%%%%%%%%%%%%%%%%%%%%%%%%%%%%%%%%%

%%%%%%%%%%%%%%%%% BODY OF PAPER %%%%%%%%%%%%%%%%%%

\section{Introduction}
Transits, whereby bodies in other systems are observed to pass in front of their host stars, have been used to great effect to explore the wealth of extrasolar planetary systems in the Galaxy \citep{FirstKeplerPaper}. The Kepler space telescope has used this technique to find over 2,600 exoplanets, some in the habitable zone, and characterise their radii and masses, discovering some of the most well known and dynamically interesting systems such as Kepler-223 \citep{Kepler-223}. Planets are not the only objects to have been detected around other stars. Transits due to smaller bodies have also been found with \citet{2018MNRAS.474.1453R} finding evidence of comets around F stars using Kepler observations of their asymmetric transits. A dust cloud released from bodies forming a tail of debris can explain both the levels and asymmetry of the transits, and enables a mass estimate of the parent bodies. \newline

One of these `dipper' stars that has so far evaded explanation, however, is the main sequence F star KIC 8462852, also known as `Boyajian's star' or `Tabby's star'. \citet{2016MNRAS.457.3988B} found, using the Kepler light curves, that the star experienced irregularly shaped transits with depths up to 20\%; these transits were aperiodic and lasted between 5 and 80 days. In addition to this a level of secular dimming was detected but the exact amount is in dispute depending on the interpretation of archival data from photographic plates \citep{2016ApJ...830L..39M,2016ApJ...822L..34S}. \citet{2016MNRAS.457.3988B} considered many possibilities for the cause of the transits but came to the conclusion that the most consistent with the data was the passage of a family of exocomets transiting at about 0.5 au. These could result from the breakup of a single body greater than 100 km in size with a minimal mass of $10^{-6} M_{\oplus}$. It has since been shown that a family of comets moving on similar orbits can reproduce the observed transits with about 700 objects with 10 km radii needed \citep{2016ApJ...819L..34B}. An alternate hypothesis was put forward by \citet{2016ApJ...829L...3W} where the transits are caused by an artificial mega-structure, known as a `Dyson sphere' or a `Dyson swarm', though this requires the presence of extra terrestrial intelligence in the system. \newline

\citet{2018MNRAS.473.5286W} extended the comet hypothesis by showing that the secular dimming could be caused by material distributed along a single elliptical orbit. Though they make no assumptions about the origin of this material, it fits well with the exocomet hypothesis where one large ($> 100$ km) body breaks up and the resultant material is spread around the progenitor's elliptical orbit. The constraints derived from the secular dimming give a transit distance between 0.05 and 0.6 au. The parent body for these comets would likely have come from a reservoir of debris left over from planet formation, like our own Kuiper belt, and was perturbed onto its current orbit. While most belts observed in other planetary systems typically exist at 10s to 100s of au from their host star, the lack of detection of an infrared excess around KIC 8462852 does not rule out a cold belt at these distances \citep{kic846prevmeas}. Given that these exocomets are inferred to transit at between 0.05 and 0.6 au from the host star, the planetesimals causing these transits must have very high eccentricities ($\sim0.99$) leaving the question: how did the parent body end up on such a highly elliptical orbit? One hypothesis originally proposed by \citet{2016MNRAS.457.3988B} is that the parent body could have evolved under action of the Kozai-Lidov mechanism. \newline

The Kozai-Lidov mechanism is a dynamical process first formulated by \citet{1962AJ.....67..591K} and \citet{1962P&SS....9..719L}. It is a three body effect that occurs when the orbital planes of two bodies orbiting the same host star are highly misaligned. The two bodies then undergo oscillations in inclination and eccentricity as they exert a gravitational torque on each other. \citet{1962AJ.....67..591K} examined this effect in the context of the perturbation of Jupiter on an inclined comet. That study neglected the effect of Jupiter's eccentricity and found that the oscillations take place for mutual inclinations $i$ in the range $\rm{cos}(\textit{i}) < 2/5$ and found a well defined relationship between the initial mutual inclination and the maximum eccentricity of the comet. In this case with a perturber on a circular orbit, the maximum eccentricity can only be appreciably large for initial mutual inclinations close to 90 degrees. Including the effects of a perturber's eccentricity leads to much more complicated behaviour; studies have shown that in this case extremely high eccentricities can be reached and the orbital plane of the perturbed body can flip from prograde to retrograde \citep{2011ApJ...742...94L}. This behaviour can occur at high inclination and low eccentricity (HiLe) or low inclination and high eccentricity (LiHe) \citep{2016ARA&A..54..441N} and is often chaotic \citep{2014ApJ...791...86L}. Though eccentricities very close to 1 can theoretically be achieved, in reality the effect of General Relativity and/or tides becomes dominant once the body gets close enough to the host star \citep{2013ApJ...773..187N}. The action of these effects is to cause a precession in the longitude of pericentre of the body's orbit which competes with that induced by the Kozai-Lidov mechanism, shutting it off if its perturbation is stronger. The dissipative effect of tides could then also act to circularise the orbit at a low pericentre and increase the timescale for the Kozai-Lidov evolution essentially decoupling the bodies from each other. Indeed, this has been proposed as a formation mechanism of both hot Jupiters and close Kuiper belt binaries \citep{KozaiKBOsTheory,KozaiKBOsObs,2012ApJ...754L..36N}. \newline

For the planetesimals in a belt around KIC 8462852 to undergo eccentricity oscillations from this mechanism, a perturber is needed. This could be an unseen planet in the system, however it would have to have become significantly inclined to the planetesimal belt at some point in its life. Planets form out of the protoplanetary disc that evolves into a debris disc once the gas has dispersed, thus it is expected that debris discs and planets should be aligned and there are many systems where this is the case including our own Solar system. However, there are planetary systems where the planets have large mutual inclinations with respect to each other like $\pi$ Men \citet{JerryPiMen}. These are thought to form from dynamical instabilities where planets undergo close encounters and scatter each other to high inclinations. Thus, it is possible for there to exist systems with high mutual inclinations between planets and debris discs (as is actually seen in HD 106906 \citep{HD106906PlanetDetection}), though close encounters that lead to inclinations high enough for the Kozai-Lidov mechanism may be highly unlikely. A more promising candidate for a misaligned perturber is a 0.4 $M_{\odot}$ M dwarf seen with small on sky separation from KIC 8462852 in Keck AO images \citep{2016MNRAS.457.3988B}. It was hypothesised to be bound as it has a similar Gaia distance estimate to KIC 8462852 of about 450 parsecs \citep{2016A&A...595A...1G}. Follow up observations by \citet{2021ApJ...909..216P} show that the two stars have the same proper motion and are in fact bound with a projected separation of 878 $\pm$ 8 au. Wide binaries such as this could potentially form through one of two pathways. The first is core fragmentation whereby the collapsing cloud of gas that the stars form from fragments into two large cores that form two stars \citep{GoodwinCoreFrag,FisherCoreFrag,OffnerCoreFrag}. The other mechanism is dynamical capture where stellar encounters within the birth cluster result in pairs of stars that formed separately becoming bound, whilst other stars are ejected, though this method is too inefficient to account for all binary stars \citep{DynamicalCapture}. Either way, it could have a random inclination to any planetesimal belt around KIC 8462852 and could potentially be highly inclined \citep{HaleIncs}, causing Kozai-Lidov oscillations of small bodies which could explain the observations. \newline

This paper aims to test how often the action of the Kozai-Lidov mechanism on a belt of planetesimals due to a wide binary companion can excite the largest planetesimals to high eccentricities. The derived occurrence rate can then be compared to the one potential detection in the Kepler field to see if the Kozai-Lidov mechanism is a likely explanation for the phenomenon. In section~\ref{section:paramexp} the parameter space of the Kozai-Lidov mechanism for an eccentric perturber is explored to investigate what orientation a general planetesimal belt has to start with to reach low pericentres and the fraction of objects that reach them. This is examined through integrating the secular equations of motion and comparing the results with N-body simulations. Section~\ref{section:MC} outlines a Monte Carlo model of binary systems in the Kepler field which is used to find the fraction of the systems that undergo Kozai-Lidov oscillations and for what fraction of their main sequence lifetimes they produce observable signatures. Section~\ref{section:results} details the results of this model for sensible system parameters, examining the most likely location of belts and companions in these systems. Section~\ref{section:discussion} illustrates the dependence of the results on the unknown parameters of the model and the choice of initial distributions as well as providing a discussion on the caveats of the model and section~\ref{section:conclusions} presents our conclusions.

\section{Parameter Space Exploration of the Eccentric Kozai-Lidov Mechanism}
\label{section:paramexp}

If we are to create a Monte Carlo model of the action of the Kozai mechanism on stars and their planetary systems in the Kepler field it is first necessary to examine how belts of planetesimals behave in the presence of an inclined companion star. Once this behaviour has been discerned, it can then be fed into the Monte Carlo model to produce a probability that the Kozai mechanism is causing the variability in the lightcurve of KIC 8462852. Specifically, the inclinations between the belt and companion star that allow low pericentres to be reached and the fraction of objects in such an inclined belt that reach a low enough `threshold' pericentre to cause observations are needed for the Monte Carlo model. \newline

\begin{figure*}
	\centering
	\includegraphics[width=\textwidth]{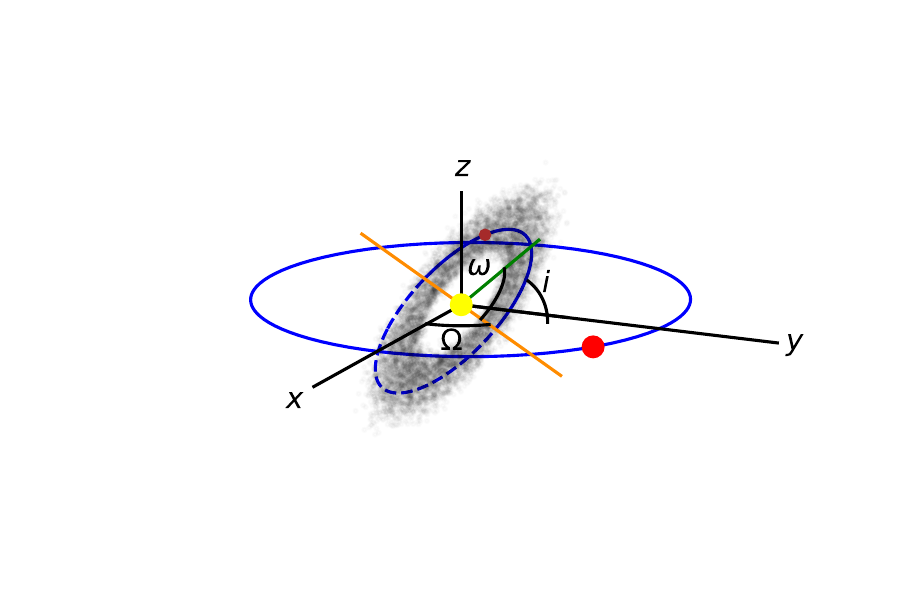}
	\caption{An example of the type of planetary system that might undergo Kozai-Lidov evolution. A host star (yellow) is orbited by a planetesimal (brown) in a belt of particles (light grey). Also in orbit around this system is a wide stellar companion (red) which has been drawn closer to the belt than expected in a hierarchical system for clarity. The inclination $i$, longitude of pericentre $\omega$ and the longitude of ascending node $\Omega$ of the planetesimal's orbit is shown to clarify their geometrical significance. The orange line represents the line of nodes, where the disc intersects the orbital plane of the companion, and the green line represents the semi-major axis of the planetesimal's orbit which is in the plane of the disc. The black lines represent a 3D coordinate basis aligned with the major, minor and perpendicular axes of the elliptical orbit of the companion. }
	\label{fig:ExampleSystem}
\end{figure*}

This work is restricted to the action of wide binary companion stars on planetesimal belts: specifically we are considering the Kozai-Lidov mechanism in the case of an external massive perturber and an internal massless perturbed object which does not exert a torque on the perturber. There are four main variables in this problem which are all orbital elements of the perturbed object as the orbital elements of the perturber do not change with time. In our context these are a planetesimal and a companion star respectively and hereafter referred to as such. The variables of the planetesimal's orbit are: the mutual inclination with respect to the companion star ($i$), the eccentricity ($e$), the longitude of ascending node as measured with respect to the plane of the binary ($\Omega$) and the longitude of pericentre ($\omega$). The basic setup of the problem is illustrated in figure \ref{fig:ExampleSystem}. Whilst these are the only variables in the problem, there are also other, constant, parameters of the system that are important. For example, the masses of the two stars contribute to the \textit{timescale} of the effect, but not its amplitude. Likewise, the semi-major axes of the two orbits and the eccentricity of the companion star affect the timescale to first order, though it has been shown that they have second order effects on the amplitude of motion \citep{2013ApJ...773..187N}. \newline

There are two ways that the variation of these orbital elements can be explored: the secular equations of motion can be integrated numerically, or N-body integrations can be used to numerically integrate Newton's second law. The latter will be more accurate but also take a prohibitive amount of time and so the full exploration of parameter space will be undertaken with the secular equations and the results compared to N-body integrations.

\subsection{The Secular Equations}

The Kozai-Lidov mechanism is a subset of the hierarchical three body problem. One comparatively massless planetesimal ($m_{\rm{pl}}$) orbits a massive host star $M_*$ which is also in a binary orbit with a companion star of mass $M_{\rm{c}}$. The system is hierarchical because $a_{\rm{pl}} \ll a_{\rm{c}}$. The Hamiltonian for the massless planetesimal $m_{\rm{pl}}$ is approximately given by 

\begin{equation}
    \label{eq:Hamiltonian}
    H^{TP} \approx \frac{3}{8} \frac{GM_* m_{\rm{pl}}}{a_{\rm{c}}} \left( \frac{a_{\rm{pl}}}{a_{\rm{c}}}\right)^2  \frac{1}{(1-e_{\rm{c}}^2)^{3/2}} (F_{\rm{quad}} + \epsilon F_{\rm{oct}}),
\end{equation}

\noindent where

\begin{equation}
    \label{eq:epsilon}
    \epsilon = \frac{a_{\rm{pl}}}{a_{\rm{c}}} \frac{e_{\rm{c}}}{1-e_{\rm{c}}^2}
\end{equation}

\noindent and $F_{\rm{quad}}$ and $F_{\rm{oct}}$ are the quadrupole and octupole contributions respectively. These are functions of the orbital elements and are listed in appendix~\ref{section:QuadOctTerms}. For all the following integrations of the secular equations we use a renormalised Hamiltonian which removes the prefactors in equation~\ref{eq:Hamiltonian}. This simply results in a renormalised time parameter $\tau$ which is related to the true time t by 

\begin{equation}
    \label{eq:renorm_time}
    t = \frac{8M_* a_{\rm{c}}^3 (1-e_{\rm{c}}^2)^{3/2}}{3m_{\rm{c}} a_{\rm{pl}}^3 \Omega_{*}} \tau,
\end{equation}

\noindent where $\Omega_*$ is the angular velocity of $m_{\rm{pl}}$ about $M_*$. \newline

\begin{figure}
    \centering
    \includegraphics[width=\columnwidth]{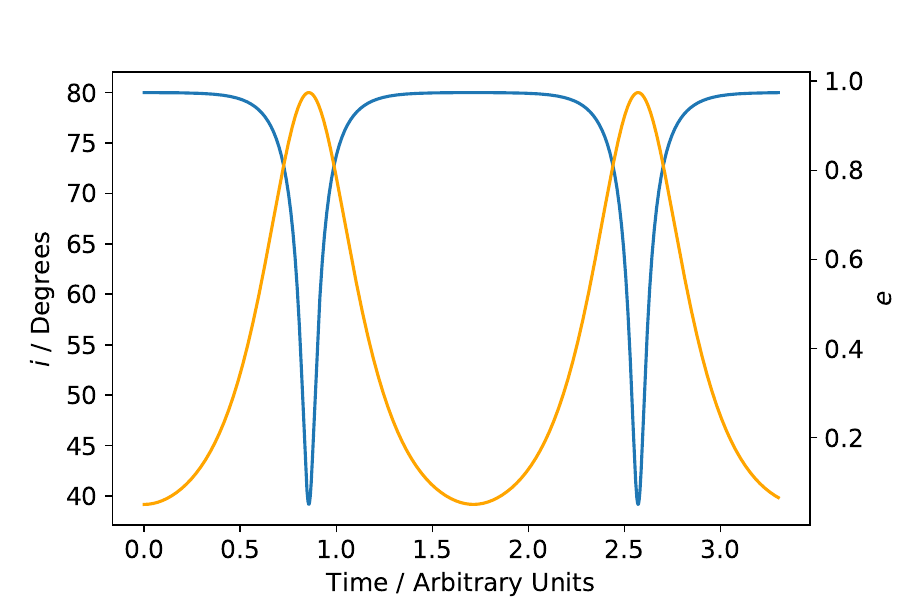}
    \caption{Evolution of the inclination (blue) and eccentricity (orange) of a test particle in the SKM scenario ($e_2 = \epsilon = 0$). The initial values of the orbital elements of the perturbed object are $i_0 = 80^{\circ}$, $e_0 = 0.05$, $\omega_0 = 180^{\circ}$ and $\Omega_0 = 0$.}
    \label{fig:SKMexample}
\end{figure}

The Hamiltonian in equation~\ref{eq:Hamiltonian} has been averaged over the longitudes of both the planetesimal and companion, expanded in the ratio $a_{\rm{pl}}/a_{\rm{c}}$ and truncated after the octupole term. This approximation is equivalent to smearing the objects out over their orbits to form a wire whose density at some point is inversely proportional to the orbital velocity at that location; these wires then exert a gravitational torque on each other proportional to their mass (so the massless planetesimal `wire' does not exert a torque on the companion). If the Hamiltonian is cut off at first order, such that only the quadrupole term is left, then the system is integrable: this is referred to as the standard Kozai-Lidov mechanism (hereafter referred to as the `SKM'). This also arises if the companion is on a circular orbit such that $e_{\rm{c}} \approx \epsilon \approx 0$. In the case of the SKM, the orbit of the planetesimal exhibits coupled oscillations in its inclination and eccentricity, becoming more eccentric and less inclined to the perturber before reversing, as illustrated in figure~\ref{fig:SKMexample}. The timescale for these oscillations to occur is given by \citep{QuadTimeEq}

\begin{equation}
\begin{split}
    \label{eq:quadtime}
    t_{\rm{quad}} = 5.3 &\left( \frac{a_{\rm{pl}}}{20 \rm{au}} \right)^{-3/2} \left( \frac{M_*}{1.43 M_{\odot}}\right)^{1/2} \left( \frac{M_{\rm{c}}}{0.4 M_{\odot}} \right)^{-1} \left( \frac{a_{\rm{c}}}{1000 \rm{au}} \right)^3 \\ &(1-e_c^2)^{3/2} \rm{Myr}.
\end{split}
\end{equation}

\noindent From this it can be seen that $t_{\rm{quad}} \gg t_{\rm{orb}}$ as the system is hierarchical. Also, the closer the planetesimal is to the companion star (i.e. the smaller the ratio $\frac{a_{\rm{c}}}{a_{\rm{pl}}}$), the faster the oscillations occur. The reason for the coupling between eccentricity and inclination in this case is because the component of the planetesimal's angular momentum that is parallel to the companion's angular momentum $J_z \propto \cos(i) \sqrt{1-e^2}$ is conserved \citep{1962AJ.....67..591K,1962P&SS....9..719L}. \newline

If the perturber has an appreciable eccentricity then $\epsilon \neq 0$ and the octupole terms in the expansion of the Hamiltonian become important; these terms can significantly change the overall dynamical behaviour of the system. Thus, the quantity $\epsilon$ acts as the `strength' of the octupole contribution and for these effects to be significant without the secular approximation breaking down it must lie in the range $10^{-3} - 10^{-1}$ \citep{2016ARA&A..54..441N}. This case is known as the eccentric Kozai-Lidov mechanism (hereafter referred to as the EKM). The timescale for these `octupole order' effects is given by

\begin{equation}
    \label{eq:octtime}
    t_{\rm{oct}} = \frac{t_{\rm{quad}}}{\epsilon^{1/2}}.
\end{equation}

Using Hamilton's equations we can find the rate of change of the planetesimal's orbital elements with time which can then be integrated numerically and thus perform a parameter space exploration. These equations are listed in appendix~\ref{section:KozaiEqs} for both the SKM ($\epsilon = 0$) and the EKM ($\epsilon \neq 0$).  \newline

\subsection{Parameter Space Exploration}
\label{section:ParamExp}
The numerical integrator used for this analysis is the LSODA package \citep{2019ascl.soft05021H,doi:10.1137/0904010}. It handles stiff and non-stiff differential equations using the BDF and Adams method respectively, automatically detecting which is needed at each timestep. The timestep it uses is variable and is set to keep the relative and absolute error tolerances below a threshold value. For all of the following work, the error tolerance is set to $10^{-11}$ in order to adequately capture the high eccentricities achieved (e $\gg$ 0).  \newline

\begin{figure*}
    \centering
	\includegraphics[width=\textwidth]{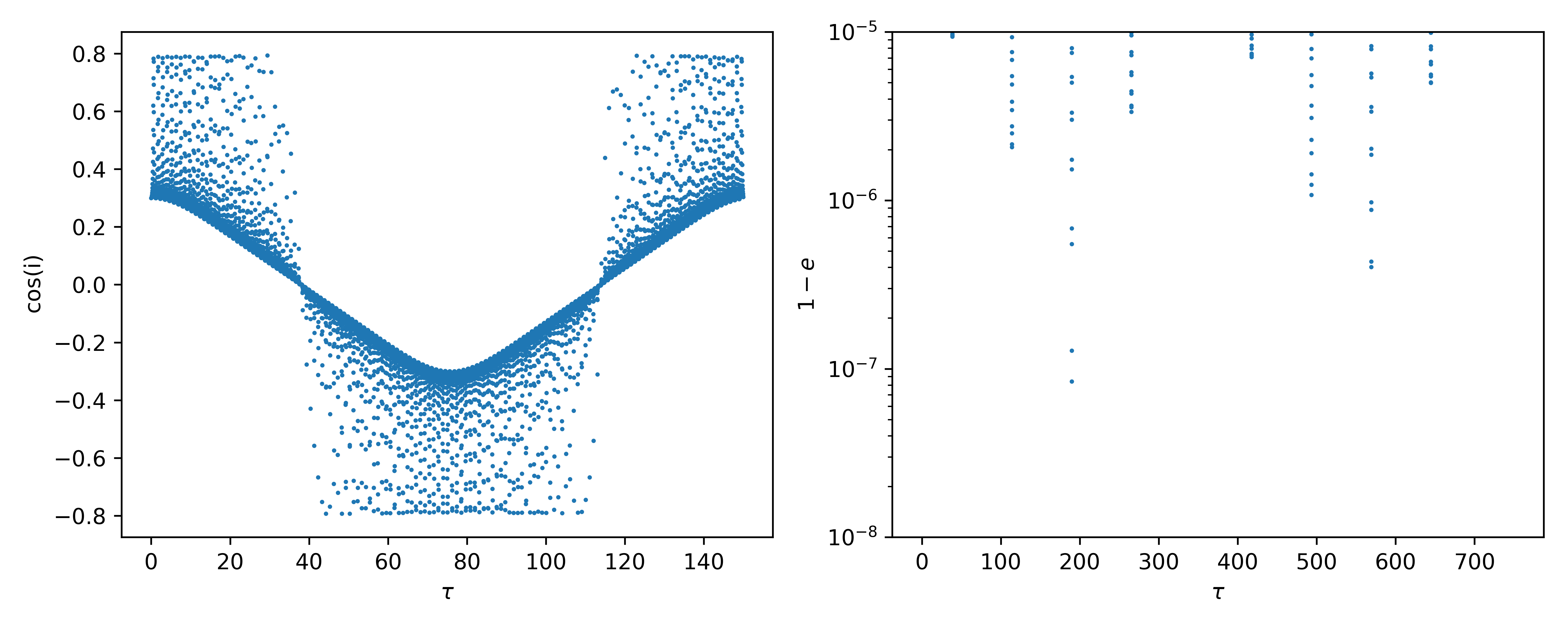}
    \caption{Evolution of a test particle in the eccentric Kozai-Lidov mechanism with $\epsilon$ = 0.01. Initial conditions were: $I_0$ = $72.5^{\circ}$, $e_0$ = 0.192, $\omega_0$ = 0, $\Omega_0$ = $\pi$. The left panel shows the time evolution of the cosine of inclination. The right panel shows the time evolution of the eccentricity, zooming in on where it gets very close to 1. Different ranges of normalised time ($\tau$) are used in each plot to highlight where the eccentricity is predicted to reach very extreme values ($1-e \approx 10^{-7}$).}
    \label{fig:EccKozExampleEvol}
\end{figure*}

The general behaviour of a particle undergoing the eccentric Kozai mechanism for a specific set of initial conditions is shown in figure~\ref{fig:EccKozExampleEvol}. The left panel shows the evolution of $\cos(i)$ for the paramters noted in the caption. The inclination oscillates on a comparatively short timescale given roughly by $t_{\rm{quad}}$ and is equivalent to evolution in the standard Kozai mechanism. This behaviour is modulated by the longer term orbital flips that happen on the comparatively longer octupole timescale $t_{\rm{oct}}$. The right panel shows the eccentricity, plotted as $1-e$, restricted to where the eccentricity is closest to 1 for clarity. It highlights the extreme eccentricities reached in this situation with the maximum being when $1 -e \sim 10^{-7}$, though it should be noted that other physical processes would prevent such a high eccentricity from ever being reached (such as GR precession, sublimation, collision with the star). This set of initial conditions was used by \citet{2011ApJ...742...94L} and these results can be compared with figures 4 and 6 from their work. \newline

\begin{figure*}
    \centering
    \includegraphics[width=\textwidth]{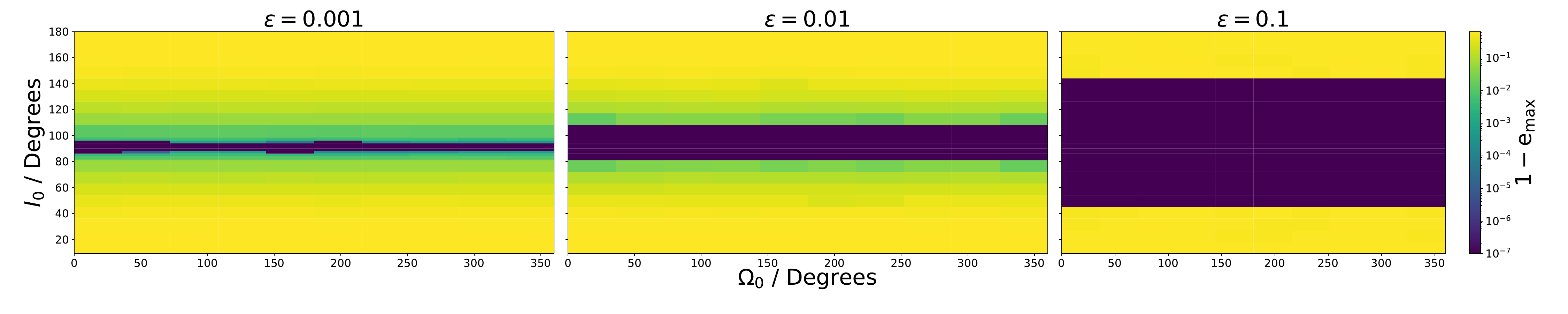}
    \caption{Minimum scaled pericentre as a function of initial inclination $i_0$ and longitude of ascending node $\Omega_0$. The colour bar shows the value of $1-e_{\rm{max}}$. To eliminate the dependence on the angles, for each value of $I_0$ and $\Omega_0$ the maximum eccentricity was calculated using 100 randomly chosen values of $\omega_0$ and a Rayleigh distribution of $e_0$ centred on 0.03. From the 100 results the maximum achievable eccentricity was taken and plotted.}
    \label{fig:Minq}
\end{figure*}

For the Monte Carlo model outlined in section~\ref{section:MC}, we will need to know the inclinations between belts and companion stars that allow planetesimals in the belt to reach high eccentricities. As we will be dealing with eccentricities very close to one, we will instead examine the `scaled pericentre' parameter defined to be

\begin{equation}
    \label{eq:scaledperidef}
    q' = 1-e_{\rm{pl}} = \frac{q_{\rm{pl}}}{a_{\rm{pl}}},
\end{equation}

which is the true pericentre of a planetesimal orbit scaled by the semi-major axis. The minimum scaled pericentre $q'$ that a planetesimal reaches will depend on its initial orbital elements: $i_0$, $\Omega_0$, $\omega_0$ and $e_0$. When considering belts of planetesimals, however, all objects in a belt will share the same initial inclination $i_0$ and longitude of ascending node $\Omega_0$ relative to a distant perturber as illustrated in figure~\ref{fig:ExampleSystem}: it is these two parameters that define the belt. Within the belt the planetesimals will have a distribution of initial eccentricities $e_0$ and longitudes of pericentre $\omega_0$. Therefore, in the context of examining how close to their host stars particles in a disc would be seen to get, it is necessary to find $\min(q'(i_0,\Omega_0;\epsilon))$. This is the minimum possible scaled pericentre that can be achieved by one of the particles in a belt defined by $i_0$ and $\Omega_0$ and is shown in figure~\ref{fig:Minq}. The value for each disc represents the minimum scaled pericentre found when doing 100 integrations with randomly distributed values of $\omega_0$ and initial eccentricites taken from a Rayleigh distribution with peak 0.03. The Rayleigh distribution of eccentricities is motivated by observations of objects in the classical Kuiper belt and from debris disc scale heights (assuming $e \sim I$) \citep{ SaiScaleHeight,YinuoScaleHeight} as well as N-body simulations of mutual planetesimal scattering \citep{IdaMakinoRayleighSims}, though in our model a companion star is perturbing the disc so the characteristic eccentricity could be higher \citep{MustillSecularStirring}. Each integration ran for a time of $\tau$ = 500 and the orbital elements were recorded at $10^7$ equally spaced intervals; the process was repeated for three values of $\epsilon=[10^{-3},10^{-2},10^{-1}]$ \newline

Figure~\ref{fig:Minq} shows that there is only a weak dependence of $\min(q(i_0,\Omega_0;\epsilon))$ on $\Omega_0$ over the probed values of $\epsilon$, whereas there is a strong dependence on $i_0$. As $\epsilon$ increases, the range of $i_0$ over which it is possible to get very low scaled pericentres increases from a small window around $90^{\circ}$ to a window that extends all the way down to $45^{\circ}$ which matches with the simulations undertaken previously by \citet{OConnorWD1856}. Figure~\ref{fig:Minq} shows that it is important to consider the EKM effects when modelling planetesimal belts in wide binaries as it widens the range of initial inclinations at which planetesimals can achieve low pericentres compared to the SKM case. In the SKM, to achieve a scaled pericentre $q'_{\rm{crit}}$, planetesimals must have initial inclinations greater than $i_{\rm{crit}}$ where 

\begin{equation}
\label{eq:SKMincrange}
    \cos(i_{\rm{crit}}) = \pm \sqrt{\frac{3}{5}(1-(1-q'_{\rm{crit}})^2)}.
\end{equation}

\noindent This leads to a `window' in initial inclination around $90^{\circ}$ within which a planetesmial will reach scaled pericentres $q'<q'_{\rm{crit}}$ and is given by 

\begin{equation}
\label{eq:window}
    \Delta i_0  \approx \frac{180}{\pi} \sqrt{\frac{24 q'_{\rm{crit}}}{5}},
\end{equation}

\noindent where $\Delta i_0$ is in degrees and $q'_{\rm{crit}} \ll 1$ is assumed. Hence, to achieve $q'_{\rm{crit}} < 10^{-4}$ in the EKM case (assuming $\epsilon = 10^{-1}$) a planetesimal must have $i_0 \geq 45^{\circ}$ as can be seen from figure~\ref{fig:Minq}, but in the SKM case, using equation~\ref{eq:window}, a planetesimal must have $i_0 \geq 88.75^{\circ}$. 

\begin{table}
    \centering
    \caption{The parameters used in the N-body integrations to achieve different octupole strengths.}
    \begin{tabular}{|c|c|c|c|}
         \hline
         $\epsilon$ & $a_1$ / au & $a_2$ / au & $e_2$  \\
         \hline
         0.001 & 88.5 & 885 & 0.01 \\
         0.01 & 87.6 & 885 & 0.1 \\
         0.1 & 18.68 & 885 & 0.9 \\
         \hline
    \end{tabular}
    \label{Tab:table}
\end{table}

\subsection{Comparison with N-body Simulations}
The validity of these results is examined with N-body integrations. The IAS15 integrator in \textsc{rebound} was used for the comparison \citep{2015MNRAS.446.1424R,2012A&A...537A.128R}. It uses a 15th order modified Runga-Kutta method and Gauss-Radau spacing and has a variable timestep to make sure the motion at pericentre is adequately captured when the orbit is highly eccentric and the particle is moving very fast. \citet{2015MNRAS.446.1424R} show that it copes well with the extreme eccentricities achieved in the EKM up to $e \sim 1-10^{-10}$ whilst maintaining an energy error of $\sim \frac{10^{-16}}{1-e_{\rm{max}}}$. \newline

\begin{figure*}
    \centering
    \includegraphics[width=\textwidth]{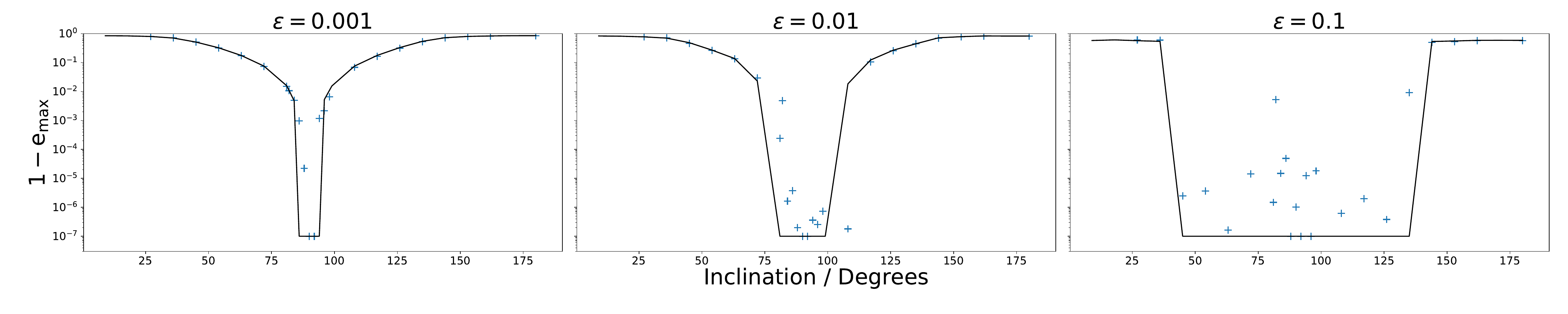}
    \caption{Comparison with N-body simulations of the maximum eccentricity for different initial inclinations. The lines are the results from integrating the secular equations and the points are found using the N-body simulations. All secular and N-body data is calculated using $\Omega=0$ and the values of $e_0$ and $\omega_0$ for the N-body points match the corresponding values used in the integration of the secular equations. All data are floored at $1-e_{\rm{max}} = 10^{-7}$ as this is the error tolerance of the integrator.}
    \label{fig:NbodyComparison}
\end{figure*}

The comparison is made with the results from integrating the secular equations and the results are plotted in figure~\ref{fig:NbodyComparison}. Simulations were run in which particles had various values of $i_0$ representing belts of different inclinations. Due to the lack of dependence of the scaled pericentre on $\Omega_0$ shown in figure~\ref{fig:Minq}, this was set to $\Omega_0=0$ for these simulations. For each N-body integration the values of $\omega_0$ and $e_0$ were chosen such that they corresponded to those that gave the lowest scaled pericentre in the secular integrations. The maximum eccentricity is then found for each simulation and compared to the same result found by integrating the secular equations. Each N-body  integration was performed three times with different particle semi-major axes and a different eccentricity of the perturber. This is done so that the evolution can be followed for $\epsilon$ values of 0.001, 0.01 and 0.1. The parameter values used to produce each octupole strength are listed in table~\ref{Tab:table} and were chosen to make sure that the test particle would not be captured by the companion due to a close approach or experience other forms of orbital evolution such as resonance \citep{2014ApJ...795..102N}. \newline

Figure~\ref{fig:NbodyComparison} shows that the results obtained when solving the secular equations agree very well with those from the N-body simulation for the case $\epsilon$ = 0.001, but that there is some disagreement with the other octupole strengths. This is probably due to the chaotic nature of the problem and the parameter space in $\omega_0$ and $e_0$ not lining up exactly between secular integrations and N-body simulations. However, we note that the general behaviour, a severe drop in scaled pericentre, is still observed for a window of inclinations around $90^{\circ}$. In fact, our work will only be interested in using scaled pericentres down to values of $10^{-4}$ and to this level the N-body simulations and secular integrations show good agreement.

\subsection{Fraction of Belt Mass Excited to High Eccentricities}
\label{section:FracExcited}
Arguably the most important parameter space exploration needed for the Monte Carlo model is the fraction of planetesimals in a belt that will reach low enough scaled pericentres to cause the transits seen in KIC 8462852 as a function of the mutual inclination between belt and companion. This is because, in the model, belts will have a wide variety of inclinations relative to their companion stars and it is therefore important to know not only whether or not it is possible for planetesimals to reach small pericentres, but also \textit{how many} of them reach these as it is not initially clear from the equations governing the secular evolution, and so we investigate it here. \newline

\begin{figure}
    \centering
    \includegraphics[width=\columnwidth]{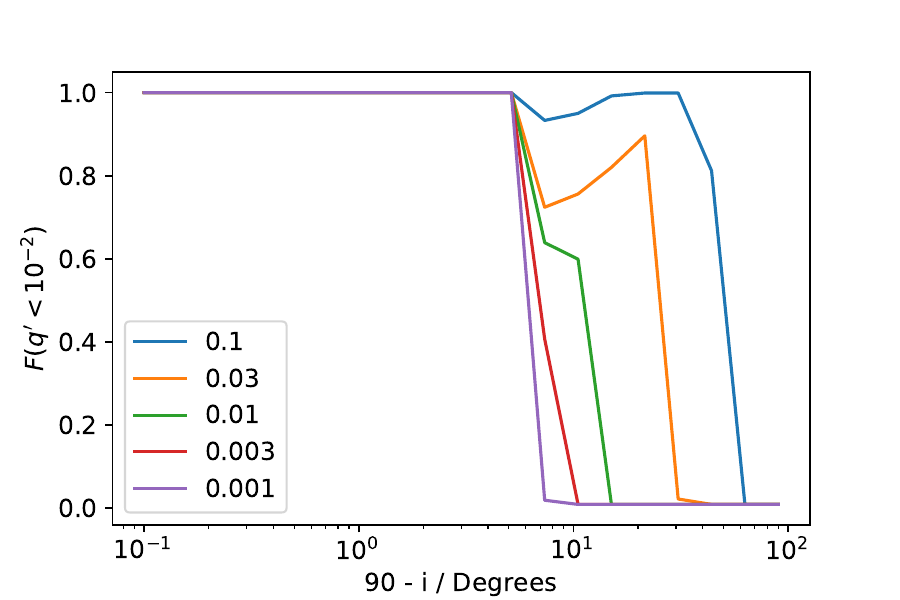}
    \caption{Fraction of planetesimal orbits that reach a scaled pericentre of at least $10^{-2}$ as a function of $i$ and $\epsilon$. Each point corresponds to 1000 integrations of orbits with uniformly distributed values of $\omega_0$ and $\Omega_0$ and a Rayleigh distribution of $e_0$ centred on 0.03.}
    \label{fig:frac1eminus2}
\end{figure}

\begin{figure}
    \centering
    \includegraphics[width=\columnwidth]{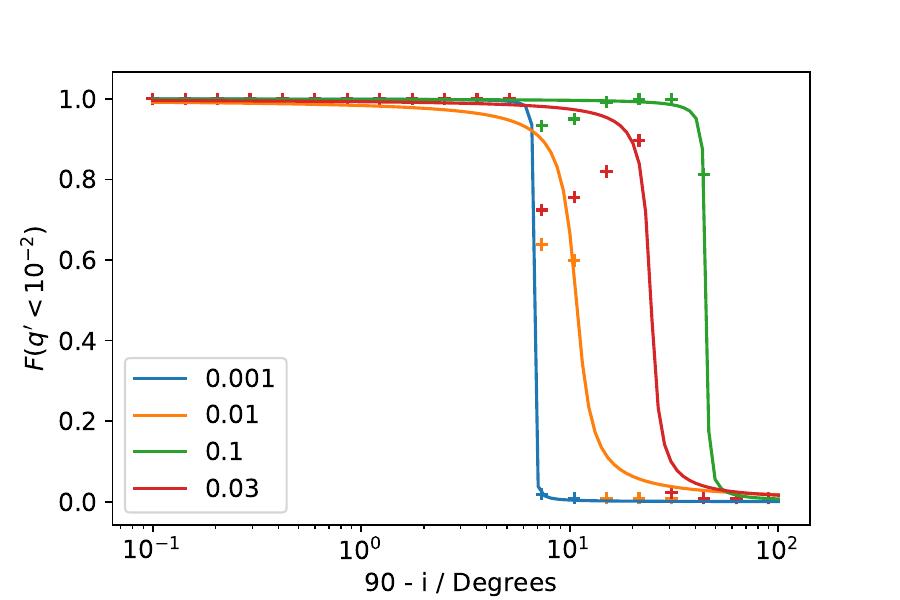}
    \caption{The fraction of randomly distributed orbits reaching a scaled pericentre less than $10^{-2}$ as a function of initial inclination for a select few values of $\epsilon$ and the fitted $\rm{tan}^{-1}$ functions as a comparison.}
    \label{fig:ArctanFits}
\end{figure}

\citet{KatzEKMEq} provide a theoretical equation that relates the inclination above which planetesimals reach `small' scaled pericentres (the level of which is undefined) $i_{\rm{crit}}$ to the value of the octupole strength $\epsilon$. From this one might theoretically assume that the fraction of objects in a belt reaching a threshold value of the scaled pericentre is a step function with its transition at $i_{\rm{crit}}$, though this is not initially obvious. In order to investigate whether this is the case, we integrate the secular equations for 1000 particles with randomly distributed values of $\Omega_0$ and $\omega_0$ and a Rayleigh distribution of eccentricities centred on 0.03. This was done for a set of inclinations that are equally spaced in $\rm{log}_{10}(90-i)$ and different values of $\epsilon$. The fraction of orbits reaching a scaled pericentre less than $10^{-2}$ (i.e., $F(q' < 10^{-2})$) is plotted in figure~\ref{fig:frac1eminus2}. The behaviour is roughly equivalent to a step function where, above some $i_{\rm{crit}}$, all objects in a belt will reach the required threshold scaled pericentre $q'_{\rm{crit}} = 10^{-2}$ and we fit the data with a $\rm{tan}^{-1}$ formula of the form 

\begin{equation}
    F(q'<10^{-2}) = \frac{1}{2} - \frac{\rm{tan}^{-1}\left(\frac{i_{\rm{mid}}-i}{\sigma}\right)}{\pi},
\end{equation}

\noindent where $i_{\rm{mid}}$ and $\sigma$ are the parameters of the fit and $i_{\rm{mid}}$ is the inclination at which 50\% of all planetesimals in the belt reach scaled pericentres less than $10^{-2}$ (i.e. $F(q'<10^{-2})=0.5$). The fits and their comparison to the data are shown for a select sample of $\epsilon$ values in figure~\ref{fig:ArctanFits}. \newline

\begin{figure}
    \centering
    \includegraphics[width=\columnwidth]{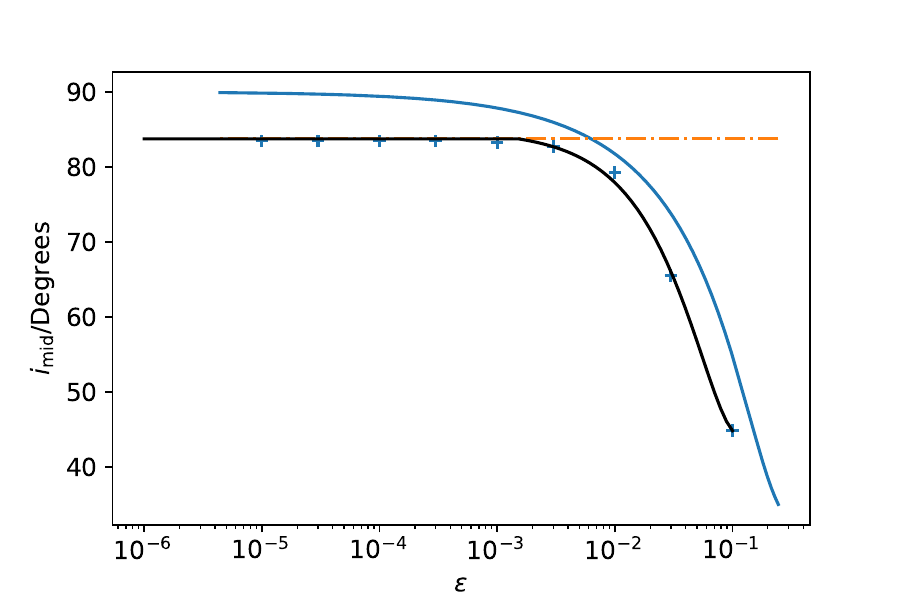}
    \caption{The values of $i_{\rm{mid}}$ from the $\rm{tan}^{-1}$ fits to the curves in figure~\ref{fig:frac1eminus2} as a function of $\epsilon$ (scatter points). The theoretical prediction from equation ref is also included for comparison (blue curve). The inclination expected when solely considering the SKM is plotted as the dot-dashed orange line. The quadratic fit to the data capped at the SKM value is shown as the black line.}
    \label{fig:imid_frac1eminus2}
\end{figure}

The values of $i_{\rm{mid}}$ for our fits are plotted as a function of $\epsilon$ in figure~\ref{fig:imid_frac1eminus2}. Comparing with the theoretical prediction from \citet{KatzEKMEq} (blue curve) shows that the equation provides the correct functional form for the dependence on $\epsilon$. However, the theoretical prediction is systematically offset towards higher inclinations which is due to the fact that this equation is not associated with a specific threshold value of the scaled pericentre, only that it is `quite small'. It is expected that, by decreasing $q'_{\rm{crit}}$ by orders of magnitude, this systematic offset would be reduced. Figure~\ref{fig:imid_frac1eminus2} also shows the value of the critical inclination needed to reach a scaled pericentre of $10^{-2}$, when solely considering the SKM case (orange dot-dashed line). This shows that, for $\epsilon < 10^{-3}$, the behaviour tends towards the standard Kozai-Lidov mechanism where the initial inclination needed to reach a maximum eccentricity of $e_{\rm{max}}$ is given simply by equation~\ref{eq:SKMincrange}. \newline

Plotted in black in figure~\ref{fig:imid_frac1eminus2} is a fit to the values of $i_{\rm{mid}}(\epsilon)$. A quadratic form is fitted, capped at the value expected from the standard Kozai-Lidov mechanism, with a best fit found to be

\begin{equation}
    i_{\rm{mid}} = A_1 \epsilon^2 + B_1 \epsilon + C_1,
\end{equation}

\noindent where $A_1 = 3237.4$, $B_1 = -723.5$ and $C_1 = 84.8$ and $\mu$ is in degrees. \newline

\begin{figure}
    \centering
    \includegraphics[width=\columnwidth]{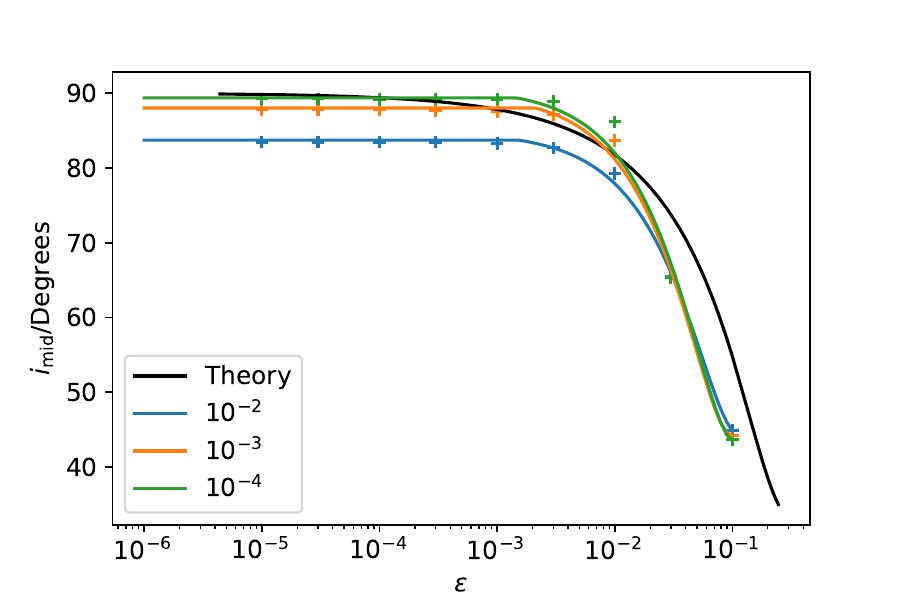}
    \caption{The best fit values of $i_{\rm{mid}}$ from fitting $\arctan$ functions to the fractions of planetesimals reaching scaled pericentres less than $10^{-2}$, $10^{-3}$ and $10^{-4}$ respectively. The black curve represents the theoretical result.}
    \label{fig:imidvsepsilon_allfracs}
\end{figure}

\noindent In addition to fitting a functional form for the fraction of planetesimals in a belt that reach a scaled pericentre of $10^{-2}$, it is necessary to examine how many planetesimals reach other, smaller scaled pericentres. This is because belt objects in the Monte Carlo model will be required to reach a \textit{physical} pericentre to produce an observational signature like Boyajian's star and as these belts will be at different radii this will translate into different scaled pericentres for each belt (see equation~\ref{eq:scaledperidef}). Figure~\ref{fig:imidvsepsilon_allfracs} illustrates the best fit values of $i_{\rm{mid}}$ for simulations where particles were required to reach scaled pericentres of $10^{-2}$, $10^{-3}$ and $10^{-4}$. The coefficients for the quadratic fit for the $10^{-3}$ and $10^{-4}$ cases are: $A_1 = 4711.3$, $B_1 = -928.9$, $C_1 = 90.0$ and $A_1 = 4443.9$, $B_1 = -916.2$, $C_1 = 90.7$ respectively.

\subsection{Summary}

In order to run a Monte Carlo model of planetesimal belts in misaligned wide binary systems it is necessary to know how the belts behave in these environments. This section has shown that, due to the EKM, large eccentricities can be reached by belt particles if the misalignment between belt and companion star is large enough. It has also shown that, when this is the case, a large fraction of belt particles reach these small scaled pericentres and has produced equations for the fraction that reach $q_{\rm{crit}}$ as a function of inclination and the octupole strength $\epsilon$.

\section{Monte Carlo Model}
\label{section:MC}
\subsection{General Setup}
\label{section:GeneralSetup}
The purpose of the Monte Carlo model is to find the expected occurrence rate of Boyajian-like stars $\langle N_{\rm{exp}} \rangle$ which are defined to be those that will have had planetary material undergo Kozai-Lidov oscillations and migrate close to the star such that they are currently producing a visible signature in the form of deep, irregular, aperiodic exocomet transits. Comparing this occurrence rate to the one system in the Kepler field will yield a probability that the `Kozai-Lidov induced eccentric exocomet' hypothesis is correct. \newline

In the model, $10^8$ stellar systems with planetesimal belts are generated, some fraction of which are binaries, whose values of the belt semi-major axis $a_{\rm{b}}$, companion semi-major axis $a_{\rm{c}}$, companion eccentricity $e_{\rm{c}}$, host star mass $M_*$ and companion star mass $M_{\rm{c}}$ are drawn from distributions such that the population will accurately reflect the Kepler field. Some fraction $f_{\rm{reject}}$ of them are rejected and cut from the sample as the EKM is prohibited from acting due to one of several physical reasons outlined in section~\ref{section:cuts}. Every system in the model is assumed to have a belt of planetesimals around each component of the binary with semi-major axis $a_{\rm{b}}$ and width $\Delta a = \frac{1}{2} a_{\rm{b}}$. The objects in the belt are assumed to undergo a collisional cascade by which larger objects collide and fragment into smaller objects and the very smallest are blown out of the system by radiation pressure. The orbits of large planetesimals in the belt are assumed to evolve due to secular interactions with the binary companion and so can, depending on the inclination of their orbit relative to the binary, migrate to small pericentres. In order to reach the roughly sub au scales associated with the transits of KIC 8462852, we require particles to achieve a pericentre less than $10^{-1} \rm{au}$ and hence a scaled pericentre less than 

\begin{equation}
 \label{eq:critscaledperi}
 q'_{\rm{crit}} < \frac{0.1}{a_{\rm{b}}/\rm{au}}.   
\end{equation}

The presence of planetesimals at these small distances could result in an observational `signature' like that for KIC 8462852 which is assumed to last for a set amount of time $t_{\rm{dur}}$, whose true value is unknown and is therefore a free parameter of the model. The fraction of the system lifetime during which this light curve signature is observable, $f_{\rm{t}}$, can be calculated for each system and the mean over all systems in the model $\overline{f_{\rm{t}}}$ can then be found. Only some of the randomly oriented planetesimals' orbits will cross the line of sight and hence have the right geometry for their dust clouds to be observationally detectable from Earth; the probability that a planetesimal's orbit causes its enveloping dust cloud of radius $R_{\rm{c}}$ to occult the stellar disc as seen from Earth is $P_{\rm{geo}}$ and is given by

\begin{equation}
\label{eq:Pgeo}
P_{\rm{geo}} = \frac{R_* + R_{\rm{c}}}{2q} \approx \frac{R_*}{2q},
\end{equation}

\noindent where it is assumed $R_{\rm{c}} \ll R_*$, $R_{\rm{c}}$ and an average has been taken over all pericentre angles \citep{WinnTransitBook}. \newline

\noindent These quantities combine to form the expected probability for a single star to be seen to undergo this behaviour

\begin{equation}
\label{eq:p}
    p = (1-f_{\rm{reject}})\overline{f_{\rm{t}}}P_{\rm{geo}},
\end{equation}

\noindent such that the expected number of stars in the Kepler field seen to exhibit this phenomenon is

\begin{equation}
\label{eq:Ntab}
    \langle N_{\rm{exp}} \rangle = 1-(1-p)^{N_{\rm{kep}}} \approx (1-f_{\rm{reject}})\overline{f_{\rm{t}}}P_{\rm{geo}} N_{\rm{Kep}},
\end{equation}

\noindent where the last relation holds if $p \ll 1$. \newline

\subsection{Finding $f_{\rm{t}}$}

The observations of KIC 8462852 are consistent with being caused by the breakup of a large $m_{\rm{crit}} \gtrsim 10^{-6} M_{\oplus}$ planetesimal. Therefore, within our model, we are only interested in the number of similar sized objects in the belt at the time small pericentres are reached $N(m>m_{\rm{crit}};t=t_{\rm{oct}})$ as they will cause transits of similar depth to KIC 8462852; the rest of the objects in the belt are ignored. The fraction of these objects $F(q'<q'_{\rm{crit}})$, that reach small enough pericentres is found using the results of section~\ref{section:FracExcited} (figure~\ref{fig:imidvsepsilon_allfracs}), where we calculate values of $F(q'<q'_{\rm{crit}})$ by interpolating between the values for $q'_{\rm{crit}} = 10^{-2}, 10^{-3}$ and $10^{-4}$. For that fraction that reach $q'_{\rm{crit}}$, they are assumed to produce an observable signature that lasts for $t_{\rm{dur}}$ Myr which is a free parameter. Hence, the total fraction of the main sequence lifetime during which transits could be observed is

\begin{equation}
\label{eq:Unsaturatedft}
f_{\rm{t}} = \frac{F(q'<q'_{\rm{crit}}) N(m>m_{\rm{crit}};t=t_{\rm{oct}}) t_{\rm{dur}}}{t_{\rm{MS}}}.
\end{equation}

However, if the system has enough bodies more massive than $m_{\rm{crit}}$ then the transits due to different objects will end up overlapping and eventually the transits will saturate. In this case the fraction of the lifetime where transits are observable is instead given by

\begin{equation}
\label{eq:Saturatedft}
f_{\rm{t}} = \frac{(t_{\rm{oct, lower}} - t_{\rm{oct, upper}}) + t_{\rm{dur}}}{t_{\rm{MS}}},
\end{equation}

\noindent where the numerator represents the range of time for which planetary material from any part of the belt will be at small pericentres. This implicitly assumes that all the material that will migrate to small pericentres will do so on the first octupole cycle and will stay there for $t_{\rm{dur}}$ until it is removed from the system. For each system in the model both the saturated and unsaturated values of $f_{\rm{t}}$ are calculated and the smaller of the two is adopted as the value for that system. \newline

\noindent As can be seen from equations~\ref{eq:Unsaturatedft} and~\ref{eq:Saturatedft}, in order to calculate $f_{\rm{t}}$, it is necessary to know the main sequence lifetime of the system. This is taken from the mass using the homology relation

\begin{equation}
\label{eq:MSlifetime}
t_{\rm{MS}} = 
\begin{cases}
    10000~M_*^{-9/2} &  \text{          for  }  M_* < 1.5~M_{\odot} \\
    3630~M_*^{-2} &  \text{          for  }  M_* > 1.5~M_{\odot}
    % $10000 M_*^{-9/2}$ &  \text{          for  }  \text{  $M_* < 1.5 M_{\odot}$} \\
    % $3630 M_*^{-2}$ &  \text{          for  }  \text{  $M_* > 1.5 M_{\odot}$}
\end{cases}
\end{equation}

\noindent where $t_{\rm{MS}}$ is in Myr and $M_*$ in $M_{\odot}$. In the saturated case it is necessary to know the octupole timescale for the belt which is given by equation~\ref{eq:octtime} but to illustrate the dependence on the orbital parameters of the problem, we rewrite it in the form given by \citet{QuadTimeEq} and used by \citet{2017MNRAS.468.4399M} as

\begin{equation}
t_{\rm{oct}} = 40 \left( \frac{M_{*}}{1.43 M_{\odot}} \right) \left( \frac{0.4 M_{\odot}}{M_{\rm{c}}} \right) \left( \frac{a_{\rm{b}}}{20 \rm{au}} \right)^{-2} \left( \frac{a_{\rm{c}}}{1000 \rm{au}} \right)^{7/2} \frac{(1-e_{\rm{c}}^2)^2}{e_{\rm{c}}^{0.5}} \rm{Myr}.
\label{eq:octtimemetzger}
\end{equation}

The timescale for planetesimals in a disc at a radius $a_{\rm{b}}$ to be excited to small enough pericentres is taken to be the value of $t_{\rm{oct}}$ at the central disc radius, however the upper and lower edges of the disc will have timescales of $t_{\rm{oct,upper}}$ and $t_{\rm{oct,lower}}$ respectively which are given by replacing $a_{\rm{b}}$ with $a_{\rm{upper}} = \frac{5}{4} a_{\rm{b}}$ and $a_{\rm{lower}} = \frac{3}{4} a_{\rm{b}}$ respectively in equation~\ref{eq:octtimemetzger}. \newline

In the unsaturated case it is necessary to know the number of particles greater than a certain mass at the time the belt undergoes the EKM $N(m>m_{\rm{crit}};t=t_{\rm{oct}})$. In order to do this the mass of the belt must be known and this requires a collisional model of the belt.

\subsection{Collisional Model}
\label{section:CollMod}

A population model for belts around main sequence sun-like stars that accounts for collisional evolution was developed by \citet{WyattDiscModel} and its free parameters were constrained by comparing with the infrared emission detected from nearby stars \citep{2018MNRAS.475.3046S}. In this model, it is assumed that all stars are born with a planetesimal belt whose masses $M_{\rm{b}}$ are drawn from a log-normal distribution centred on $M_{\rm{mid}}$ which is a free parameter. These belts orbit a host star of mass $M_*$ at semi-major axis $a_{\rm{b}}$ and have a blackbody radius $R_{\rm{bb}}$, drawn from a power law distribution with exponent $\gamma$ within the range $1 < R_{\rm{bb}}/\rm{au} < 1000$, i.e. 

\begin{equation} 
\label{eq:PRbb}
P(R_{\rm{bb}}) \propto 
\begin{cases}
R_{\rm{bb}}^{\gamma} & 1 < R_{\rm{bb}}/\rm{au} < 1000 \\
0 & \textrm{Otherwise.}
\end{cases}
\end{equation}

In this model these belts are assumed to undergo collisional evolution where large bodies that have been stirred onto crossing orbits will collide and catastrophically disrupt to form smaller bodies. The planetesimals have a diameter D which varies between the maximum size $D_{\rm{c}}$ which is set by planet formation processes when the system is born, and the blowout size $D_{\rm{bl}}$ at which radiation pressure puts dust grains onto unbound orbits. Planetesimals in the belt are assumed to have a size distribution of the form 

\begin{equation}
    \label{eq:SizeDist}
    n(D) = KD^{-\alpha},
\end{equation}

\noindent where $\alpha$ is 3.5 in an infinite collisional cascade \citep{InfiniteCascadePowerLaw} and K is a normalisation constant. Assuming that the mass is the only significant time variable quantity, then the disc mass evolves according to

\begin{equation}
    M = \frac{M(0)}{1 + t/t_{\rm{c}}(0)},
\end{equation}

\noindent where $t_{\rm{c}}(0)$ is the initial collisional timescale of the largest bodies in the belt. Assuming that particles have a Rayleigh distribution of eccentricities with means $\langle e \rangle = \langle i \rangle$, and that the fractional size of an object that will catastrophically destroy a planetesimal $X_{\rm{c}} \ll 1$, \citet{WyattDiscModel} find that the mass of a disc at times $t_{\rm{age}} >> t_{\rm{c}}(0)$ is given by

\begin{equation}
    M = \num{1.4e-9} r^{13/3} (dr/r) D_{\rm{c}} {Q_{\rm{D}}^{*}}^{5/6} \langle e \rangle ^{-5/3} M_*^{-4/3} t_{\rm{age}}^{-1},
\end{equation}

\noindent {where $Q^*_{\rm{D}}$ is the dispersal threshold of a planetesimal, $\langle e \rangle$ is the peak of the distribution of eccentricities, $dr$ is the width of the belt, $D_{\rm{c}}$ is the maximum size of planetesimal and $t_{\rm{age}}$ is the age of the system. This can be expressed more simply as

\begin{equation}
\label{eq:BeltMassWithUnknownParams}
    M = M_*^{-4/3} t_{\rm{age}}^{-1} r^{13/3} M_{\rm{mid}} A/B,
\end{equation}

\noindent where $A = D_{\rm{c}}^{1/2} {Q_{\rm{D}}^*}^{5/6} e^{-5/3}$ and $B = D_{\rm{c}}^{-1/2} M_{\rm{mid}}$. A similar equation that also depends on A and B can be found for the fractional luminosity of these discs (assuming black body emission) and the population model was compared to observations of fractional excesses of nearby systems by \citet{2018MNRAS.475.3046S}. This enabled best fit values for the parameters A, B and $\gamma$ which could be well constrained, albeit with some degeneracy, since varying $B$ changes the initial fractional luminosity distribution that belts are born with and varying $A$ changes the fractional luminosity distribution at late times. \citet{2018MNRAS.475.3046S} find best fit values of $A = \num{5.5e5} \rm{km}^{1/2} \rm{J}^{5/6} \rm{kg}^{-5/6}$, $B = 0.1 M_{\oplus} \rm{km}^{-1/2}$ and $\gamma = -1.7$ and these values of A and B are used in the equation for the masses of our belts~\ref{eq:BeltMassWithUnknownParams} and the value $\gamma$ is the exponent in our power law distribution of belt radii. \newline

Using the model of \citet{WyattDiscModel} with the above best fit values, the corresponding total mass in the belt $M_{\rm{bb}}$ at the time when the EKM excites planetesimals to small pericentres, $t_{\rm{oct}}$, can be found

\begin{equation}
    \label{eq:Mblackbody}
    M_{\rm{bb}}=  \num{1.75e-7} R_{\rm{bb}}^{13/3} t_{\rm{oct}}^{-1} M_{\rm{mid}},
\end{equation}

\noindent where $M_{\rm{bb}}$ and $M_{\rm{mid}}$ are in $M_{\oplus}$, $R_{\rm{bb}}$ is the black body radius of the belt in au and $t_{\rm{oct}}$ is in Myr. The population model of \citet{2018MNRAS.475.3046S} was fitted to the distribution of infrared excesses of nearby stars and hence constrain the distribution of temperatures of discs in the population which are assumed to emit like the blackbody of the temperature appropriate for their radius. This is why the blackbody radius is used in equation~\ref{eq:Mblackbody} and the belt mass is correct assuming blackbody emission. However, since dust grains emit inefficiently in a manner dependent on their size and composition \citep{Krivov2006}, discs are hotter than expected for their radius which means the distribution of these disc radii is likely to be different to that of their black body radius. \cite{2015MNRAS.454.3207P} found that the blackbody radius of a debris disc $R_{\rm{bb}}$ derived from fitting SEDs does not exactly match the physical radius from resolved millimetre images $R_{\rm{mm}}$, which we identify with $a_{\rm{b}}$, but instead differs by a factor $R_{\rm{mm}} = \Gamma R_{\rm{bb}}$ which depends on the luminosity of the disc hosting star. As the belt's radius is increased by a factor of $\Gamma$, the mass must be increased by a factor of $\Gamma^2$ in order to maintain the same distribution of fractional luminosities. This is equivalent to the argument that the cross-sectional area has decreased by a factor $Q^{-1}$ where $Q$ is the absorption efficiency of dust particles averaged over the dust temperature, this is assumed to be constant and is equivalent to $\Gamma^{-2}$. Thus, the true maximum mass of belts in this model is given by

\begin{equation}
    \label{eq:MmaxTabby}
    M_{\rm{max}}=  \num{1.75e-7} \Gamma^2 R_{\rm{bb}}^{13/3} t_{\rm{oct}}^{-1} M_{\rm{mid}}.
\end{equation}

The most recent analysis shows that the best fitting  functional form of $\Gamma$ is given by \citep{Pawellek2021}

\begin{equation}
\label{eq:gamma}
    \Gamma = 2.92\left( \frac{L_*}{L_{\odot}} \right)^{-0.13},
\end{equation}

\noindent and, for this model, we follow the methodology of \citet{Pearce2022} which uses equation~\ref{eq:gamma}, capped at a maximum value of 4, to convert $R_{\rm{bb}}$ to $R_{\rm{mm}}$. However, in order to make use of equation~\ref{eq:gamma}, the luminosity of each star in the sample must be known and hence it is assumed that the sample stars follow the power law Mass-Luminosity relation given in \citet{Eker2015MLRelation} and expanded upon in \citet{Eker2018ExpandedMLRelation}. \newline

Equation~\ref{eq:MmaxTabby} is only valid at $t \gg t_{\rm{coll}}$, i.e. at times greater than the collisional lifetime of objects in the belt. At earlier times, the belt has not begun to collisionally deplete and no small dust has been produced and blown out of the system by radiation pressure. Thus, at these early times, belts will retain their initial mass $M_{\rm{init}}=M_{\rm{mid}}$ and so we adopt the following formalism for the mass of belts at a time $t_{\rm{oct}}$

\begin{equation}
    \label{eq:Mbelt}
    M_{\rm{b}} = min(M_{\rm{max}},M_{\rm{mid}}).
\end{equation}

Using this formalism for the mass of the belt, the number of objects with masses greater than $m_{\rm{crit}}$ at $t_{\rm{oct}}$, $N(m>m_{\rm{crit}};t=t_{\rm{oct}})$, can be found. Using equation~\ref{eq:SizeDist} we can write the number of objects per unit belt mass with a mass between $m$ and $m+dm$ as

\begin{equation}
    \label{sizedistperbeltmass}
    \frac{n(m)}{M_{\rm{b}}} = \frac{1}{6}  m_{\rm{max}}^{-1/6}  m^{-11/6},
\end{equation}

\noindent where $m_{\rm{max}}$ is the mass of the largest object of diameter $D_{\rm{max}}$. Integrating this expression we find that the number of objects with a mass greater than $m_{\rm{crit}}$ per unit belt mass ($n'_{\rm{c}}$) is 

\begin{equation}
    \label{eq:ngreatermcrit}
    n'_{\rm{c}} = \frac{N(m>m_{\rm{crit}})}{M_{\rm{b}}} = \frac{1}{5} \left[(m_{\rm{max}}m_{\rm{crit}}^{5})^{-1/6} - m_{\rm{max}}^{-1} \right],
\end{equation}

\noindent where $m_{\rm{crit}}$ and $m_{\rm{max}}$ are in $M_{\oplus}$. \newline

\subsection{Incorporating the Collisional Model}
Now that we have a collisional model for the belt mass we can return to our formalism for $f_{\rm{t}}$ and elucidate its dependence on the physical variables of the system and the different regimes it can lie in. Taking the simpler case, in the saturated regime, we can substitute equation \ref{eq:octtimemetzger} into equation \ref{eq:Saturatedft} replacing $a_{\rm{b}}$ with $a_{\rm{b}} + \Delta a_{\rm{b}}/2$ and $a_{\rm{b}} - \Delta a_{\rm{b}}/2$ for $t_{\rm{oct,upper}}$ and $t_{\rm{oct,lower}}$ respectively. This leads to the following equation for $f_{\rm{t}}$

\begin{equation}
	f_{\rm{t}} = \frac{1280}{9t_{\rm{MS}}} \left(\frac{M_*}{1.43}\right) \left(\frac{0.4}{M_{\rm{c}}}\right) \left(\frac{a_{\rm{c}}}{1000}\right)^{7/2} \frac{(1-e_{\rm{c}}^2)^2}{e_{\rm{c}}^{1/2}} \left(\frac{a_{\rm{b,mid}}}{20}\right)^{-2} + \frac{t_{\rm{dur}}}{t_{\rm{MS}}}.
	\label{eq:ftsat}
\end{equation}

In the unsaturated case, assuming the belt mass has not been capped at its upper limit of $M_{\rm{mid}}$, we can substitute equations \ref{eq:MmaxTabby} and \ref{eq:ngreatermcrit} into equation \ref{eq:Unsaturatedft}. This yields the following for $f_{\rm{t}}$

\begin{equation}
\begin{split}
	f_{\rm{t}} = \num{1.1e-10} &\frac{n’_{\rm{c}} t_{\rm{dur}}  \Gamma^{-7/3} F(q'<q'_{\rm{crit}})}{t_{\rm{MS}}} \left(\frac{1.43}{M_*}\right) \left(\frac{M_{\rm{c}}}{0.4}\right) \\ &\left(\frac{a_{\rm{c}}}{1000}\right)^{-7/2} \frac{e_{\rm{c}}^{1/2}}{(1-e_{\rm{c}}^2)^2} a_{\rm{b,mid}}^{19/3}.
	\label{eq:ftunsat}
\end{split}
\end{equation}

\subsection{Input Distributions}

Having developed a model that will calculate the expected number of Boyajian-like stars in the Kepler field, it is important that the distributions of the input parameters also match observations to give a realistic output. Section~\ref{section:stellarmassinput} contains the stellar mass distribution, section~\ref{section:BeltSMAInput} the belt radius distribution, section~\ref{section:BinarySMA} the binary semi-major axis distribution and~\ref{section:BinaryEcc} the binary eccentricity distribution.

\subsubsection{Stellar Masses}
\label{section:stellarmassinput}
One important property of stars in the model is their mass, since both the timescale for the EKM interaction and the main sequence lifetime of the system depend on it, both of which affect $f_{\rm{t}}$. Higher mass stars have much shorter lifetimes than lower mass stars so there will be less opportunity for their discs to undergo Kozai-Lidov oscillations before the stars end their lives, though those that do spend a greater fraction of their lifetime doing so than an equivalent lower mass star. In order to compare our results with the Kepler field we use the observed mass distribution for this set of $\sim 200,000$ stars. The mass distribution of the Kepler field from which the masses of the primary stars, $M_*$, are drawn is shown in figure~\ref{fig:MassDist}. For the secondary stars, we instead draw masses, $M_{\rm{c}}$, from a random distribution between values of 0 and $M_*$ for each binary pair. Figure~\ref{fig:MassDist} shows the resultant total mass distribution which is different from that of the Kepler field. Though different, the primary star masses follow the Kepler distribution and secondary stars are mostly sub-solar M-dwarfs which might not have been resolved or detected by Kepler (as was the case for KIC 8462852). It is possible to make the total distribution of masses which is identical to the Kepler distribution by picking both primary and secondary masses from such a distribution, but this does not produce a uniform distribution of mass ratios nor is it consistent with observations of binary stars \citep{2010ApJS..190....1R}.

\begin{figure}
    \centering
    \includegraphics[width=\columnwidth]{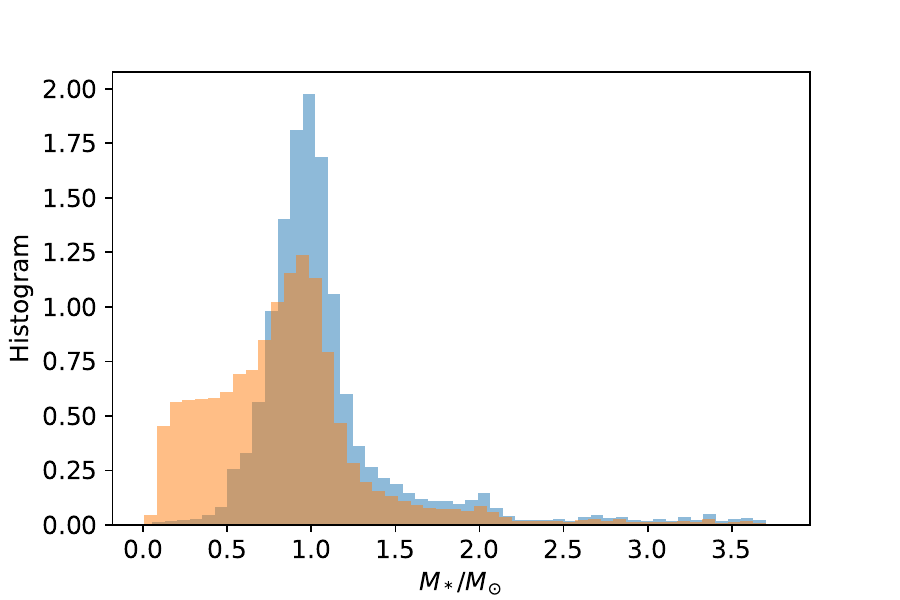}
    \caption{Mass distribution of the primary stars (blue) and the total sample including secondaries (orange). It is a combination of the distribution of the Kepler field plus lower mass companions that correspond to unobserved M-dwarfs like that of the companion of KIC 8462852. The lowest mass star is $0.086 M_{\odot}$ and the highest mass is $3.7 M_{\odot}$}
    \label{fig:MassDist}
\end{figure}

\subsubsection{Belt Semi-Major Axis}
\label{section:BeltSMAInput}
Planetesimal belts can have a range of radii as can be seen from our own system, with belts at $\sim 3$ au and $\sim 30$ au, whilst exoplanetary systems have been found to host belts that are quite massive and can extend to hundreds of au \citep{DEBRIS} and this range must be incorporated into the model. As an equation for the mass of belts was used from \citet{2018MNRAS.475.3046S} which assumed a power law distribution of debris disc radii, the same radius distribution must also be used here for consistency. The power law exponent (equation~\ref{eq:PRbb}), whose best fit value was found to be -1.7, cannot be altered without also altering the best fit values for A and B in equation~\ref{eq:BeltMassWithUnknownParams} in a consistent manner which is beyond the scope of this work. \newline

\noindent The best fit value of this exponent is such that there are more belts at small radii than large, this is because lots of belts at small radii were needed in \citet{2018MNRAS.475.3046S} to account for the fact that only 20\% of stars had an infrared excess. As every star was assumed to host a belt in this analysis, most of the population had to have close-in belts that would collisionally deplete fast enough such that most stars would have no detectable excess from a belt and this is reflected in the initial distribution of blackbody radii shown in figure~\ref{fig:abelthist}

\begin{figure}
	\centering
	\includegraphics[width=\columnwidth]{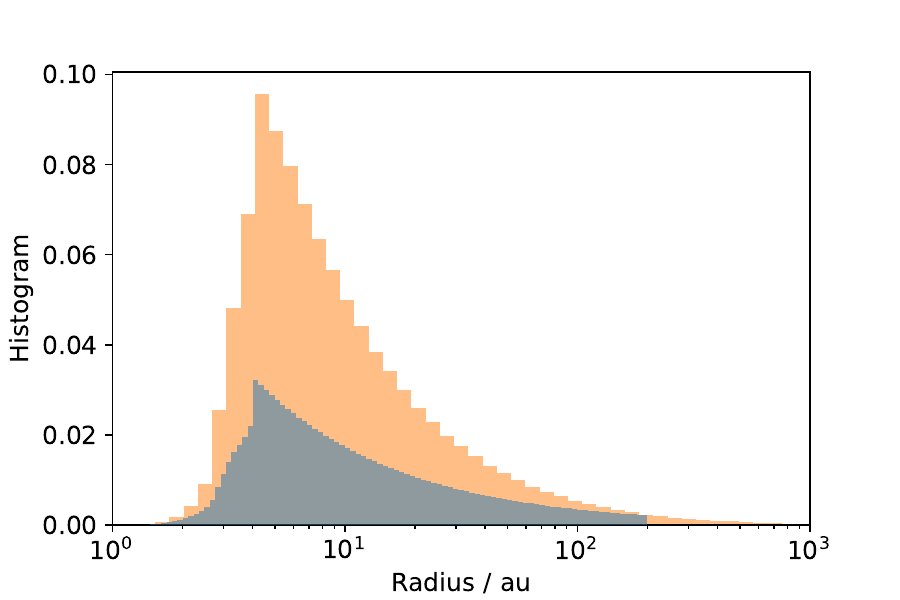}
	\caption{Histogram of the initial belt radii $a_{\rm{b}}$ in the model (blue) and those that remain after imposing the cuts outlined in section~\ref{section:cuts} (orange). The discontinuity in the pre-cut sample is due to the conversion between the blackbody radii and true radii as the conversion factor is capped at a maximum of 4 (see equation~\ref{eq:gamma} and  section~\ref{section:CollMod}).}
	\label{fig:abelthist}
\end{figure}

\subsubsection{Wide Binary Semi-Major axis}
\label{section:BinarySMA}
Around 50\% of solar-like stars in the local galaxy are gravitationally bound to other stars \citep{DucheneReview,MoeBinaries,Duquennoy,2010ApJS..190....1R}. The most common configuration is a binary pair which have a wide distribution of possible semi-major axes that can be wide (100s to 1000s of au) or close (1-10s of au), though higher order hierarchical systems such as triples and quadruples also exist. Despite the obvious hindrance of the gravitational pull of a second body, multiple systems seem to be remarkably resilient when it comes to planet formation. Planets have been found both orbiting both stars in a close pair (P-type/circumbinary), e.g. Kepler-16 \citep{Kep16} and also around one star in wide pair (S-type/wide binary planet) e.g. Kepler-444A \citep{Kep444}. In addition to planets, planet-forming discs have also been detected around binary stars \citep{CircumbinaryDiscs}. Therefore, it can be expected that, especially in wide binary systems where the star is far away and its perturbation smaller, planetesimal belts will still exist around each star. Indeed, studies have shown that planetesimal belt formation is only suppressed by intermediate binaries (10s to 100s au) \citep{MediumSepBinariesLackDDs}. \newline

In the Monte Carlo model it is assumed that 30\% of the stars are binaries and, of those that are, a log-normal period distribution centred on $10^5$ days is used as found observationally by \citet{2010ApJS..190....1R}. This period distribution is combined with the mass distribution described in section~\ref{section:stellarmassinput} to give the semi-major axis distribution shown in blue in figure~\ref{fig:acomphist}.

\begin{figure}
	\centering
	\includegraphics[width=\columnwidth]{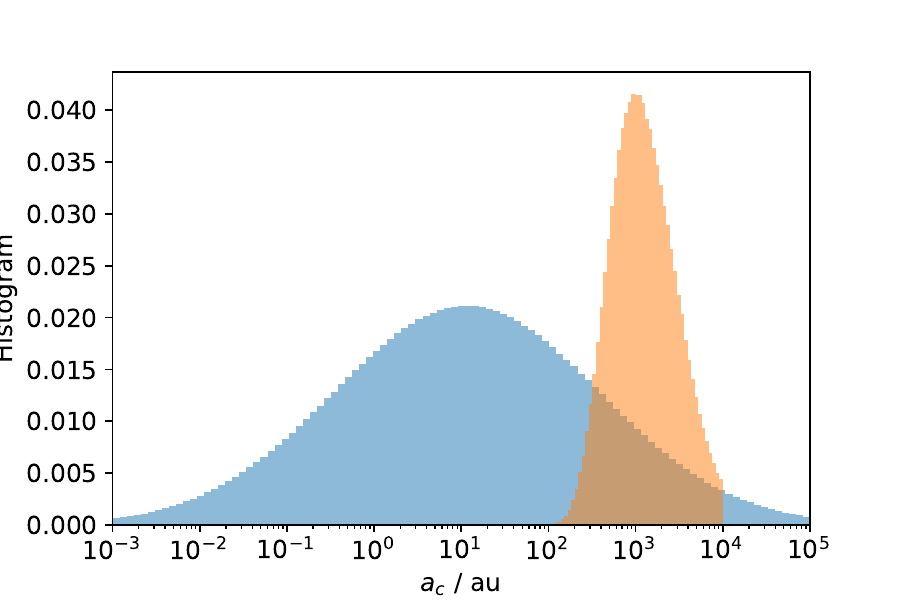}
	\caption{Histogram of the initial companion semi-major axes $a_{\rm{c}}$ in the model (blue) and those that remain after imposing the cuts outlined in section~\ref{section:cuts} (orange).}
	\label{fig:acomphist}
\end{figure}

\subsubsection{Wide Binary Eccentricity}
\label{section:BinaryEcc}
Wide binaries are thought to form through core fragmentation or dynamical capture and, due to the nature of these formation mechanisms, a wide distribution of eccentricities is expected \citep{Corefrag}. There are currently two competing interpretations of the data on wide binary eccentricities: that they have a thermal distribution where $P(e) \propto e$ \citep{ThermalEccDist} or a uniform distribution as argued for by \citet{2010ApJS..190....1R}. Although surveys of the widest binaries are biased against the highest eccentricities, in order to be consistent with the sourcing of the semi-major axis distribution from \citet{2010ApJS..190....1R}, we adopt the uniform eccentricity distribution in our model but check that the results do not change significantly when using a thermal distribution.

\subsection{Cuts to Initial Distribution}
\label{section:cuts}
In order to analyse the Monte Carlo model effectively, it is important to identify and remove systems where our setup is incompatible with a belt of particles undergoing Kozai-Lidov oscillations. These systems can then be cut from the model to leave only those that are capable of this behaviour which will allow us to see the most likely locations of belts and companions that are experiencing this effect. There are many reasons why a system might not be able to undergo Kozai-Lidov oscillations and the specific reasons examined here are: the companion star is too close to the belt and causes chaotic motion of disc particles (section~\ref{section:starbeltsep}), the companion star's orbital period is comparable to the timescale for secular evolution thus invalidating the equations of motion (section~\ref{section:starbeltsep}), the belt is too close to its host star such that GR effects shut off the Kozai-Lidov mechanism (section~\ref{section:GR}), the stars leave the main sequence before objects reach small pericentres (section~\ref{section:agecut}), and the lack of any companion star at all (section~\ref{section:binfrac}). The combined effect of these cuts is to reject a fraction $f_{\rm{reject}} = 0.986$ of the initial systems in the model.

\subsubsection{Star-Belt Separation}
\label{section:starbeltsep}
Not all separations between a companion star and a planetesimal belt will lead to Kozai-Lidov oscillations. The mechanism is hierarchical in nature, so systems where the companion star is too close to the belt will not experience this effect. The peak of initial values of $a_{\rm{c}}$ as shown in figure~\ref{fig:acomphist} is located at $\sim 10$ au. The distribution of $a_{\rm{b}}$, meanwhile, shows closer in belts are more common ($\sim 1$ au). However, there is some overlap of far-out belts with close-in companions and these are not nearly hierarchical enough for the EKM to take effect. This is not to say that particles will not reach very small pericentres through some other mechanism, secular chaos or scattering for example \citep{SecularChaos,KirkwoodGaps}, however this Monte Carlo model has been set up to specifically examine the EKM effect due to wide binary companions and thus any system that cannot undergo this phenomenon is excluded. This will include all systems where $a_{\rm{c}} < a_{\rm{b,upper}}$, i.e. where the companion star is within the belt and where the belt is outside the star (e.g. a P-type binary) and $53\%$ of the systems satisfy this condition. Also excluded is the case where the star is outside the belt but sufficiently close to expose the disc particles to chaotic evolution. The formula for the semi-major axis below which this occurs is given by equation 1 in \citet{ChaosBoundary}, this is proportional to $a_{\rm{c}}$ with the proportionality factor depending only on the eccentricity of the companion star and the masses of both bodies. $24\%$ of all systems in the model have belts located in the chaotic zone of their companions. \newline

Further to this, the K-L mechanism is a secular effect and this approximation requires that the timescale for the secular effect is greater than the orbital periods of the bodies in the system; this translates to the requirement that the smallest secular timescale $t_{\rm{quad}}$ be much larger than the largest orbital period $t_{\rm{orb,comp}}$ and for this analysis we cut any system where $t_{\rm{quad}} < 10 t_{\rm{orb,comp}}$ which corresponds to $18.9\%$ of systems. \newline

The cut on the secular timescales imposes a relation between the variables of the model that will bound the results of later calculations. Using equation \ref{eq:quadtime} for $t_{\rm{quad}}$ and $t_{\rm{orb,comp}} = 10^{-6} \sqrt{\frac{a_{\rm{c}}^3}{M_*}} \rm{Myr}$, then by requiring $\frac{t_{\rm{orb,comp}}}{t_{\rm{quad}}} = 10^{-1}$ we get the relationship at the boundary of the cut

\begin{equation}
    \label{eq:secularcut}
    a_{\rm{c}} \approx 0.158 \left(\frac{M_{\rm{c}}}{M_*}\right)^{2/3} (1-e_{\rm{c}}^2)^{-1} a_{\rm{b}}.
\end{equation}

\subsubsection{General Relativity}
\label{section:GR}

The effect of General Relativity is to induce a pericentre precession in any planetesimals which increases in strength closer to the host star; if this is stronger than the precession due to the EKM, it will dominate and the EKM will not manifest. The strength of general relativistic effects can be approximated in Newtonian gravity as a perturbation term that falls off with distance as $r^{-3}$, thus only belts that are sufficiently close to their host stars, and with sufficiently distant companions, will experience this shut off. \citet{GRFormalism} derive an $\epsilon_{\rm{GR}}$ analogous to that for the EKM given by

\begin{equation}
    \label{eq:epsgr}
    \epsilon_{\rm{GR}} = B \frac{a_{\rm{c}}^3 M_*^2}{a_{\rm{b}}^4 M_{\rm{c}}},
\end{equation}

\noindent where B is $\num{1e-8}$ such that masses are in $M_{\odot}$ and semi-major axes are in au. We can then impose the cut $\epsilon_{\rm{GR}} < 1$ such that Kozai-Lidov evolution is not shut off by General Relativity. This cut removes the systems with the closest belts and the furthest companions and $4.8\%$ of the initial sample violates this criterion. Using equation \ref{eq:epsgr} and requiring $\epsilon_{\rm{GR}} = 1$ at the boundary of the cut, we can obtain the following relation between the parameters of the systems at this boundary

\begin{equation}
    \label{eq:GRcut}
    a_{\rm{c}} = \left( \frac{1}{B} \frac{M_{\rm{c}}}{M_*^2} \right)^{1/3} a_{\rm{b}}^{4/3}.
\end{equation}

However, this analysis only excludes discs whose precession due to GR is greater than that of the EKM in their initial low eccentricity state and hence will not deviate from a belt structure at all. There will be some belts in the model where this is not true and the particles in these belts will begin to evolve to higher eccentricities. However, the pericentre precession due to GR depends on the pericentre distance, $q$, as well as the semi-major axis and hence the precession rate due to GR will increase during their evolution and eventually eclipse that of the Kozai mechanism. While the particles in these belts reach high eccentricities, some of them may not meet the threshold eccentricities to start producing strange Boyajian star-like light curves before GR takes over (i.e they do not reach $q'<q'_{\rm{crit}}$) and these systems must also be rejected from the sample. To do this we use equation 51 from \citet{GR_Treatment_Liu} which gives the minimum scaled pericentre achievable due to GR, $q'_{\rm{min,GR}}$, as

\begin{equation}
    \sqrt{q'_{\rm{min,GR}}(2-q'_{\rm{min,GR}})} = \frac{1}{9} \left(  4 \epsilon_{\rm{GR}} + \sqrt{16 \epsilon^2_{\rm{GR}} + 135 \cos^2(i_0)} \right)
\end{equation}

\noindent and those systems which cannot achieve the required scaled pericentre (i.e. $q'_{\rm{min,GR}}>q'_{\rm{crit}}$) are removed from the model.

\subsubsection{System Age}
\label{section:agecut}
The octupole timescales of the systems initially drawn from our distributions, given by equation~\ref{eq:octtimemetzger}, span many orders of magnitude. The systems with a calculated $t_{\rm{oct}}$ that is implausibly small are removed by the cut that requires the orbital timescale to be much smaller than the secular timescale. The systems with $t_{\rm{oct}}$ so large that they would never undergo Kozai-Lidov evolution in the lifetime of the universe also get removed from the model as they fall within the GR cut. These cuts still leave a variety of octupole timescales ranging from $10^5 - 10^{12}$ years. We exclude systems that do not undergo Kozai-Lidov oscillations before the star turns off the main sequence and evolves into a white dwarf as we want to compare with observations of main sequence stars in the Kepler field. Thus we require that $t_{\rm{oct}} \leq t_{\rm{MS}}$ and $4.5\%$ of the initial systems violate this criteion. This imposes another relation between the system parameters at the boundary of the cut which can be found by setting $t_{\rm{oct}}$ (given by equation \ref{eq:octtimemetzger}) equal to $t_{\rm{MS}}$ and is given by

\begin{equation}
    \label{eq:agecut}
    a_{\rm{c}} = 1000 \left(  \frac{t_{\rm{MS}}}{16000} \left( \frac{1.43}{M_*} \right) \left( \frac{M_{\rm{c}}}{0.4} \right) \frac{e_{\rm{c}}^{1/2}}{(1-e_{\rm{c}}^2)^2}  \right)^{2/7} a_{\rm{b}}^{4/7}.
\end{equation}

We also remove all systems whose octupole timescales are smaller that 10 Myr; this is because at earlier times the system is still in its planet formation stage and has a protoplanetary disc. Studies have shown that the action of the Kozai-Lidov mechanism on such a disk causes eccentric gas and dust ring formation \citep{KozaiGas}. However, it is unclear if any massive and highly eccentric planetesimals that are uncoupled to the gas would be able to produce a KIC 8462852-like signature given the surrounding gas will have a non-negligible optical depth. As this scenario is uncertain, we exclude it from our analysis. $88\%$ of systems in the model have octupole timescales shorter than 10 Myr and thus violate this cut.

\subsubsection{Binarity Fraction} 
\label{section:binfrac}
As evidenced by our own solar system, not every star is in a binary pair and hence the fraction of stars that are in binaries needs to be included. Stellar surveys show that the general binarity fraction for FGK stars that dominate the Kepler sample is about $30\%$ (see \citet{DucheneReview} and references therein). Imposing this final cut, along with all the previous cuts from sections~\ref{section:starbeltsep},~\ref{section:GR} and~\ref{section:agecut} leads to 98.6\% of all initial systems in the model being removed, leaving only $1.4\%$ of the initial systems to undergo Kozai-Lidov oscillations if they have the correct orientation.

\section{Results}
\label{section:results}
The main output of the Monte Carlo model is $\overline{f_{\rm{t}}}$ which is the mean value of the fraction of the main sequence lifetime that a system spends with large objects at small pericentres causing an observational signature and is found to be $\overline{f_{\rm{t}}} = \num{2.7e-4}$ for $t_{\rm{dur}} = 100$ yr. This value is a mean over the entire sample and sections~\ref{section:ft_abelt_acomp},~\ref{section:ft_mstar} and~\ref{section:ft_age} will elucidate its origin with respect to the main parameters of the model: $a_{\rm{b}}$, $a_{\rm{c}}$, $M_*$ and $t_{\rm{oct}}$. Unless otherwise stated, all calculations and plots assume $t_{\rm{dur}} = 100$ yr.

\subsection{Dependence of $f_{\rm{t}}$ on semi-major axes}
\label{section:ft_abelt_acomp}
The two most consequential parameters in the model are $a_{\rm{b}}$ and $a_{\rm{c}}$. Figure~\ref{fig:abeltvsacomp2D} shows the number of systems that survive the cuts of section~\ref{section:cuts} and illustrates the effect of these cuts and the belt and companion parameters that can potentially cause exocomet transits via the EKM. It shows that the majority of the systems have close-in belts with $a_{\rm{b}} < 10$ au and companion separations between $100 \lesssim a_{\rm{c}} \lesssim 3000$ au. As expected, companions with large belt radius $a_{\rm{b}} \sim 100$ au but small companion separation $a_{\rm{c}} \sim 100-1000$ au are removed due to the secular timescale $t_{\rm{quad}}$ being too similar to the orbital timescale of the companion $t_{\rm{orb}}$. As can be seen from equation \ref{eq:secularcut}, this translates to a lower bound on $a_{\rm{c}}$ of the form $a_{\rm{c}} \propto a_{\rm{b}}$ which is seen sculpting the lower edge of the population in figure~\ref{fig:abeltvsacomp2D}. Similarly, close-in belts ($a_{\rm{b}} \sim 1-10$ au) and distant companions $a_{\rm{c}} \sim 1000-10000$ au are removed because the precession due to GR is greater than that of the Kozai-Lidov mechanism. This imposes another lower bound of the form $a_{\rm{c}} \propto a_{\rm{b}}^{4/3}$ and this can clearly be seen in figure~\ref{fig:abeltvsacomp2D} plotted as the red bounding line. \newline

\begin{figure}
    \centering
    \includegraphics[width=\columnwidth]{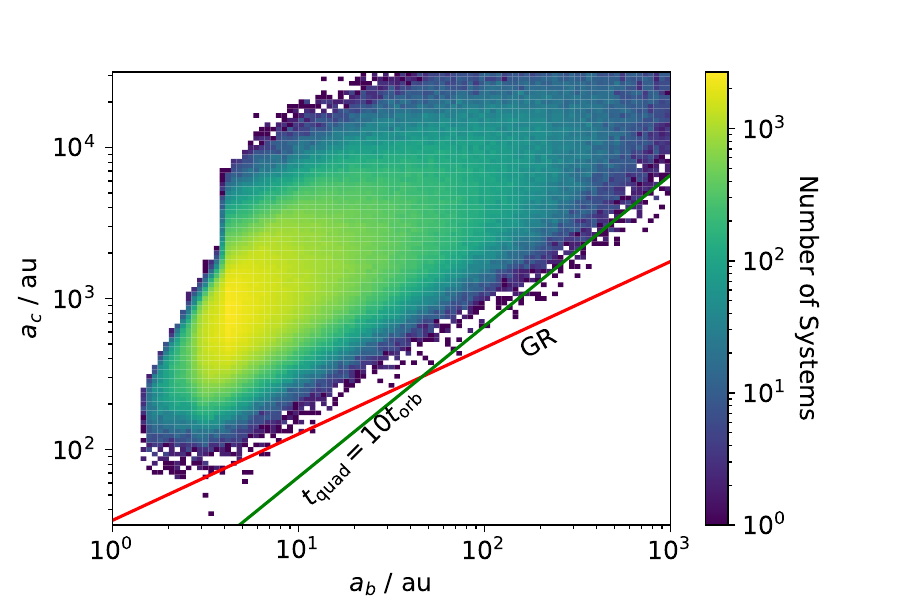}
    \caption{2D histogram of the belt radii and companion semi-major axis for every the systems that survived all the cuts. The orange line shows the boundary of the parameter space due to GR, found by substituting equation~\ref{eq:GRcut} into~\ref{eq:ftunsat} and using $t_{\rm{MS}} = 10$ Myr, $e_{\rm{c}} = 0.1$, $M_{\rm{c}} = 1 M_{\odot}$ and $M_{*} = 1 M_{\odot}$. The green line shows the boundary of the parameter space due to the timescale for secular quadrupole oscillations being 10 companion orbital timescales using $M_* = 3 M_{\odot}$, $M_{\rm{c}} = 0.1 M_{\odot}$ and $e_{\rm{c}} = 0.1$. The artefact at $a_{\rm{b}} = 4$ au is due to a majority (but not all) of the systems having a true radii that are the maximum of 4 blackbody radii according to the prescription laid out in section~\ref{section:CollMod}}
    \label{fig:abeltvsacomp2D}
\end{figure}

In order to understand where the mean value of $f_{\rm{t}}$ comes from, it is important to first examine how it depends on the variables of the model. Figure \ref{fig:abeltvsft2D} shows how $f_{\rm{t}}$ depends on the belt radius $a_{\rm{b}}$ for the belts expected to undergo EKM. The dominant relation seen in the figure is given by $f_{\rm{t}} \propto a_{\rm{b}}^{19/3}$ and arises from equation~\ref{eq:ftunsat} as most systems are in the unsaturated regime. It shows that the furthest belts spent most of their life transiting, a direct result of the longer collisional lifetime, and hence larger masses, of more distant belts at the time they undergo EKM. The upper bound of this behaviour (plotted as the upper red line in figure~\ref{fig:abeltvsft2D}) is set merely by the lifetime of the system and the cuts made to the initial population have very little effect.  \newline

\begin{figure}
    \centering
    \includegraphics[width=\columnwidth]{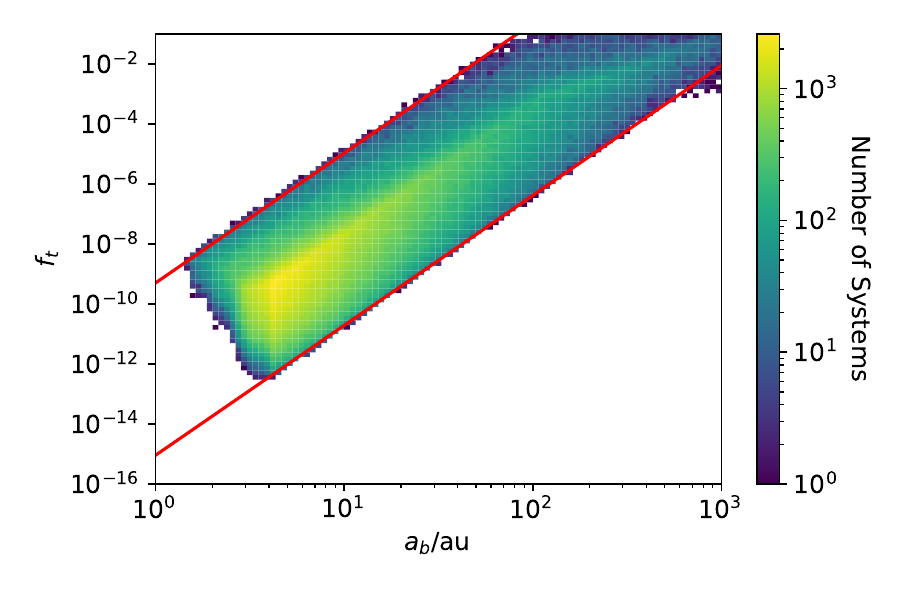}
    \caption{2D histogram of $f_{\rm{t}}$ and $a_{\rm{b}}$ for every system that survived all the cuts. The red lines indicate the relation between $f_{\rm{t}}$ and $a_{\rm{b}}$ given by equation~\ref{eq:ftunsat} using appropriate values for the other parameters of the system. The upper bounding line uses $t_{\rm{MS}} = 200$ Myr, $a_{\rm{c}} = 100$ au, $e_{\rm{comp}} = 0.1$ and $M_* = M_{\rm{c}} = 1 M_{\odot}$. The lower bounding line uses $t_{\rm{MS}} = 10^4$ Myr, $a_{\rm{c}} = 2000$ au, $e_{\rm{comp}} = 0.9$ and $M_* = M_{\rm{c}} = 1 M_{\odot}$.}
    \label{fig:abeltvsft2D}
\end{figure}

Figure \ref{fig:acompvsft2D} shows how $f_{\rm{t}}$ depends on the companion semi-major axis $a_{\rm{c}}$. Naively, it might be expected that the relationship between $f_{\rm{t}}$ and $a_{\rm{c}}$ would be given by $f_{\rm{t}} \propto a_{\rm{c}}^{-7/2}$ as this is what is given by equation~\ref{eq:ftunsat} which gave the correct relation between $f_{\rm{t}}$ and the belt radius. This relation can indeed be seen bounding the lower region of the parameter space in figure~\ref{fig:acompvsft2D} as the negatively sloped line. However, the dominant relation between $f_{\rm{t}}$ and $a_{\rm{c}}$ is given instead by $f_{\rm{t}} \propto a_{\rm{c}}^{91/12}$ such that $f_{\rm{t}}$ increases with companion semi-major axis. This is not expected from equation~\ref{eq:ftunsat} as more distant companions should take longer to destabilise belts which would then have lost mass through collisions. This result is instead due to the cut discussed in section~\ref{section:agecut}, where the EKM timescale must be less than the main sequence lifetime ($t_{\rm{oct}} < t_{\rm{MS}}$). This leads to the relation between $a_{\rm{c}} \propto a_{\rm{b}}^{4/7}$ along the boundary of the cut as seen in equation~\ref{eq:agecut} which, substituting into equation~\ref{eq:ftunsat}, gives us the relation $f_{\rm{t}} \propto a_{\rm{c}}^{91/12}$ that is seen bounding the upper and lower regions of the parameter space in figure~\ref{fig:acompvsft2D}. The second lower bound that is the most important below $a_{\rm{c}} \sim 10^4$ au is due to the requirement that the secular timescale be much longer than the orbital timescales as laid out in section~\ref{section:starbeltsep}. As shown by equation~\ref{eq:secularcut}, this leads the relation $a_{\rm{b}} \propto a_{\rm{c}}$ along the boundary and, substituting this into equation~\ref{eq:ftunsat}, generates the  observed relation $f_{\rm{t}} \propto a_{\rm{b}}^{17/6}$ at the lower edge. Hence, the overall effect of all the cuts made to the initial population is that the fraction of time a system will spend with large objects at small pericentres actually \textit{increases} with $a_{\rm{c}}$ rather than decreasing. \newline

\begin{figure}
    \centering
    \includegraphics[width=\columnwidth]{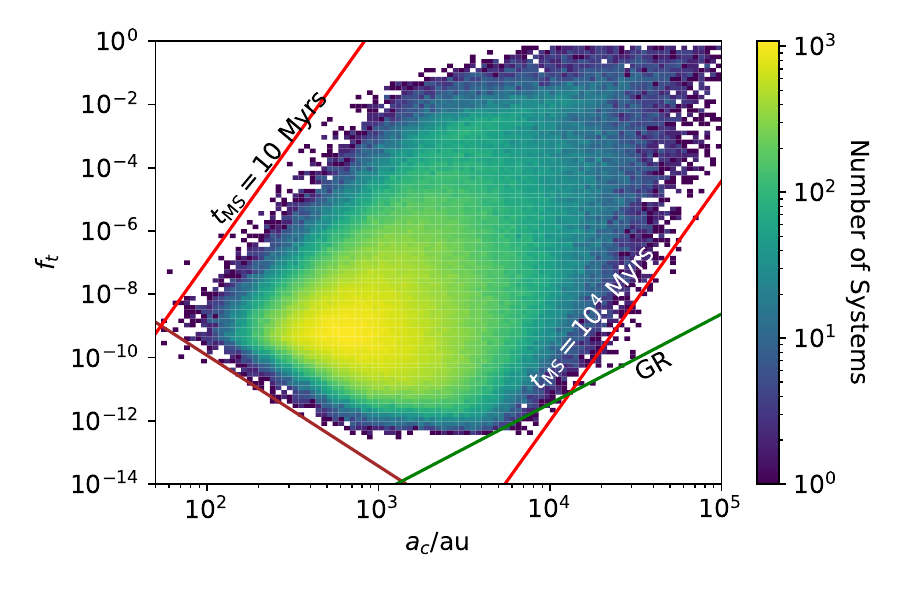}
    \caption{2D histogram of $f_{\rm{t}}$ and $a_{\rm{c}}$ for every system that survived all the cuts. The brown line indicates the relation between $f_{\rm{t}}$ and $a_{\rm{c}}$ given by equation~\ref{eq:ftunsat} using $t_{\rm{MS}} = 1400$ Myr, $a_{\rm{b}} = 1$ au, $e_{\rm{comp}} = 0.1$ and $M_* = 1 M_{\odot}$ and $M_{\rm{c}} = 0.1 M_{\odot}$. The two red lines represent the boundary of the parameter space outside of which $t_{\rm{MS}}>t_{\rm{EKM}}$ whose limit is given by equation~\ref{eq:agecut}. The upper red line assumes $t_{\rm{MS}} = 10^4$ Myr, $e_{\rm{comp}} = 0.9$ and $M_* = M_{\rm{c}} = 1 M_{\odot}$. The lower red line assumes $t_{\rm{MS}} = 10$ Myr, $e_{\rm{comp}} = 0.1$ and $M_* = 1.43 M_{\odot}$ and $M_{\rm{c}} = 1 M_{\odot}$. The green line represents the edge of the parameter space below which GR shuts off the EKM whose boundary is given by equation~\ref{eq:GRcut} and assumes $t_{\rm{MS}} = 10^4$ Myr, $e_{\rm{comp}} = 0.9$ and $M_* = 1 M_{\odot}$ and $M_{\rm{c}} = 0.2 M_{\odot}$.}
    \label{fig:acompvsft2D}
\end{figure}

Figures \ref{fig:abeltvsft2D} and \ref{fig:acompvsft2D} show that systems with more distant belts (up to $10^3$ au) and more distant companion stars ($\sim 10000$ au) have the largest values of $f_{\rm{t}}$ and hence spend the greatest fraction of their main sequence lifetime in the `transiting' state. However, this does not account for the rarity of these systems. Indeed figures~\ref{fig:abelthist} and~\ref{fig:acomphist} show that most systems have close-in belts ($\sim 1-10 \rm{au}$) and close companions ($\sim 100-1000 \rm{au}$). These most common systems spend much less of their lifetime in the transiting state and hence skew the mean value of $f_{\rm{t}}$ to lower values. \newline

\begin{figure}
    \centering
    \includegraphics[width=\columnwidth]{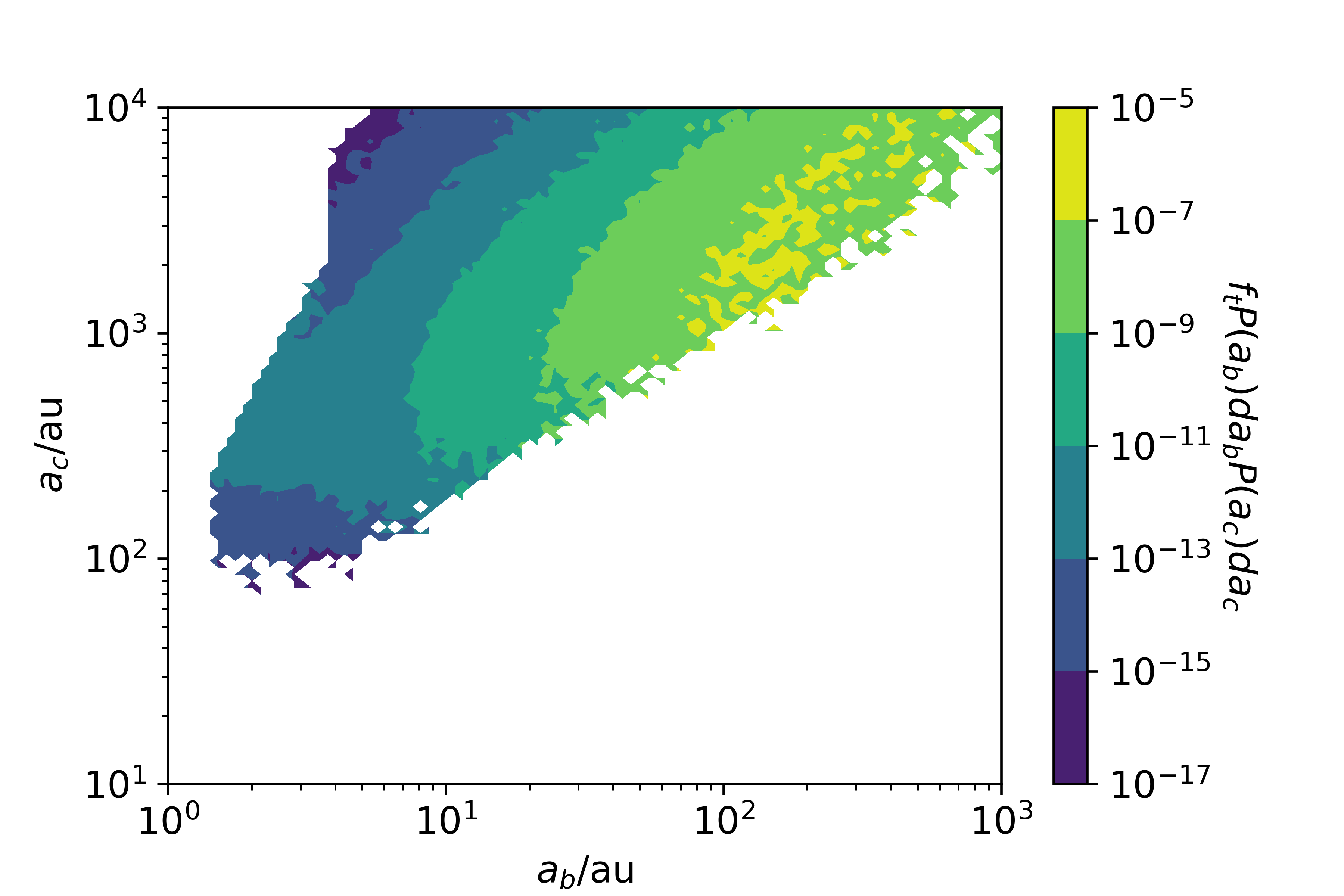}
    \caption{Mean value of $f_{\rm{t}}$ multiplied by the probability for a system to be in that bin as a function of belt radius $a_{\rm{b}}$ and companion semi-major axis $a_{\rm{c}}$ such that the sum of the values at each point gives the mean $f_{\rm{t}}$ over all systems in the model. The bin size is $0.03 \rm{dex}^2$ and the artefact at $a_{b} = 4$ au is due to a majority (but not all) of the systems having a true radii that are the maximum of 4 blackbody radii according to the prescription laid out in section~\ref{section:CollMod}. }
    \label{fig:BeltandCompLoc}
\end{figure}

It is important, however, to find the most likely systems to be observed, and the greatest contributors to $\overline{f_{\rm{t}}}$. Figure~\ref{fig:BeltandCompLoc} shows $f_{\rm{t}}P(a_{\rm{c}})da_{\rm{c}}P(a_{\rm{b}})da_{\rm{b}}$ which is the local mean of $f_{\rm{t}}$ in $a_{\rm{b}}$ and $a_{\rm{c}}$, multiplied by the probability distributions of those parameters. The distributions used are those of the post-cut population shown in orange in figures~\ref{fig:abelthist} and~\ref{fig:acomphist}. It can be seen that the most likely systems to be seen transiting, and that dominate the contribution to the mean value, are those that have belts in the range 100-1000 au and companions in the range 300-10000 au.

\subsection{Dependence of $f_{\rm{t}}$ on Stellar Mass}
\label{section:ft_mstar}

\begin{figure}
    \centering
    \includegraphics[width=\columnwidth]{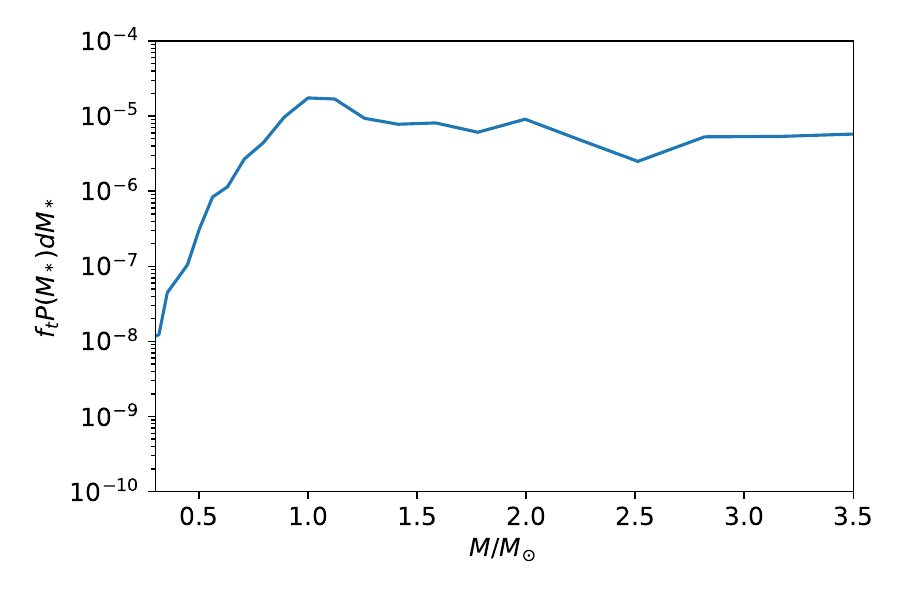}
    \caption{Average value of $f_{\rm{t}}$ as a function of $M_{*}$, the primary star mass, weighted by the Kepler mass probability density function.}
    \label{fig:mstarvsft}
\end{figure}

Figure~\ref{fig:mstarvsft} illustrates how $f_{\rm{t}}$ depends on the mass of the stars in the system. In an analogous manner to figure~\ref{fig:BeltandCompLoc}, it shows $f_{\rm{t}}(M_*)P(M_*)dM_*$ which is the local mean of $f_{\rm{t}}$ in stellar mass multiplied by the stellar mass probability distribution. The latter is taken to be the mass distribution of stars observed by Kepler (fig~\ref{fig:MassDist}) rather than the expected stellar mass function of the Galactic field in order to match the results to the Kepler field. Whilst more massive host stars undergo Kozai-Lidov oscillations more slowly (equation~\ref{eq:octtimemetzger}) and hence do not have many large objects left by that time, they also have a much shorter lifetime: hence $f_{\rm{t}}$ is larger for these systems. The reverse is true for less massive host stars, whilst they have more massive belts at the time of Kozai-Lidov, they have much longer lifetimes and hence are less likely to be observed with large objects at small pericentres. This increasing trend with stellar mass persists despite the high bias towards solar mass stars in the Kepler field, though the increase levels off after 1 solar mass.

\subsection{Dependence of $f_{\rm{t}}$ on system age}
\label{section:ft_age}

Figure~\ref{fig:agevsft} shows the dependence of $f_{\rm{t}}$ on the octupole timescale of the system $t_{\rm{oct}}$, weighted by the probability distribution of octupole timescales. As $t_{\rm{oct}}$ represents when systems would first excite large objects to small pericentres, this is roughly equivalent to the age of the system when the observable signatures of cometary transits would become visible in the lightcurves of these stars. It shows that the most likely systems to exhibit this phenomenon are stars that are roughly $10^2-10^3$ Myrs old, whilst below $10^2$ Myrs and there is a downturn. The downturn below $10^2$ Myrs is only slight, however, before it reaches the stage where systems would still be in the protoplanetary disc phase ($t_{\rm{oct}} \sim 10$ Myr) below which systems are cut from the model. Above $\sim 10^3$ Myrs, systems become less likely to be observed in a transiting state and this is due to a combination of factors. Firstly, from figure~\ref{fig:mstarvsft}, more massive stars are more likely to be seen to transit due to their shorter lifetimes, hence stars are unlikely to be seen transiting at $\sim 10$ Gyr ages as all the high mass stars have left the main sequence and the low mass stars will either have transit events earlier on in their lives or $t_{\rm{oct}}$ is $\sim 10$ Gyr long but the belt has been severely depleted.

\begin{figure}
    \centering
    \includegraphics[width=\columnwidth]{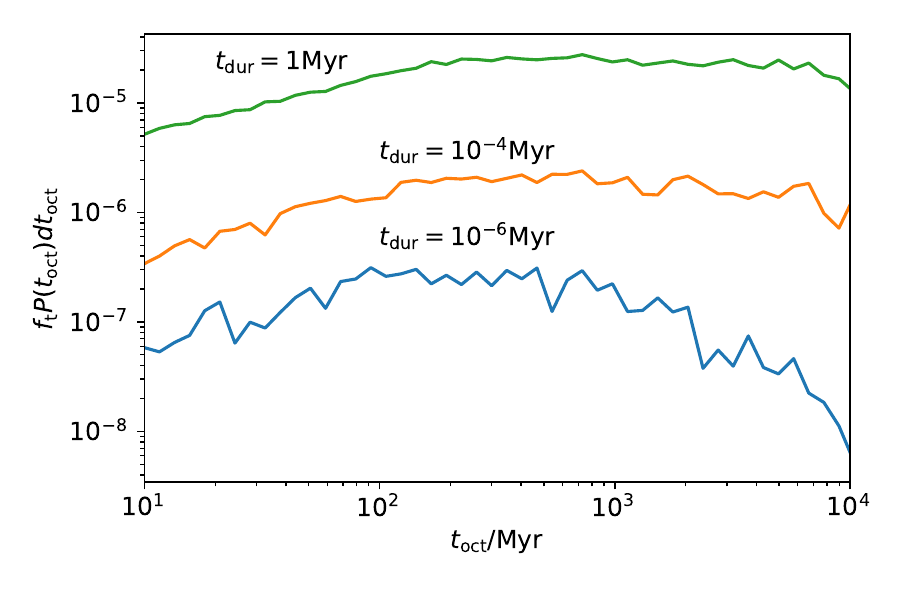}
    \caption{Mean fraction of lifetime spent with large objects at small pericentres as a function of the octupole timescale $t_{\rm{oct}}$ and weighted by the probability distribution of $t_{\rm{oct}}$. This timescale roughly corresponds to the stellar age at the time when transits would become observable and hence shows what age stars that exhibit cometary lightcurves would be expected to be.}
    \label{fig:agevsft}
\end{figure}

\subsection{Probability of the EKM as the cause of observations}
\label{section:NexpAndft}
The mean fraction of their lifetime that stars in the Kepler field spend with large planetesimals at scaled pericentres $q'<10^{-2}$ is found to be $\overline{f_{\rm{t}}} = \num{2.7e-4}$. In order to turn this into an expected number of KIC 8462852-like objects in the Kepler field ($N_{\rm{exp}}$) we first use equation~\ref{eq:p} to find the probability an individual star exhibits KIC 8462852-like dips. Using the homology relation $R_* \propto M_*^{1/13}$ and $q = 0.6$ au from the observations of KIC 8462852, a value of $P_{\rm{geo}}$ for each star can be found which, due to the weak dependence of $R_*$ on $M_*$, varies little from system to system and has a mean value of $\overline{P}_{\rm{geo}} =  \num{3.8e-3}$. Combining $\overline{f_{\rm{t}}}$ with $\overline{P}_{\rm{geo}}$ and $f_{\rm{reject}}$ yields $p = \num{6.6e-9}$ and, as $p \ll 1$, equation~\ref{eq:Ntab} gives the probability of observing one or more stars to undergo these KIC 8462852-like dimming events in the Kepler field as $\langle N_{\rm{exp}} \rangle = \num{1.3e-3}$.  \newline

This can also be framed in a Bayesian sense. If the occurrence rate of stars with a KIC 8462852-like lightcurve P(L) is 1/200,000 from Kepler observations, and the occurrence rate of said stars if their properties are due to comet scattering via the Kozai mechanism P(L|K) is $\overline{f_{\rm{t}}} (1-f_{\rm{reject}}) \overline{P}_{\rm{geo}} = \num{6.6e-9}$, then using Bayes' theorem the probability of the Kozai mechanism causing the strange lightcurve observations P(K|L) is:

\begin{equation}
    P(K|L) = \frac{P(L|K)P(K)}{P(L)} = \num{1.3e-3},
\end{equation}

\noindent where it is assumed that P(K), the probability that the Kozai mechanism will take effect in the systems, disregarding the considerations already made, is unity. \newline

Figure~\ref{fig:fthist} shows the distribution of non-zero values of $f_{\rm{t}}$ in the sample of the $\sim 1\%$ of systems that were not rejected and shows that the majority of the values of $f_{\rm{t}}$ sit below the mean. The fact that the majority of systems spend a very small fraction of their lifetime in the transiting stage is to be expected. This is chiefly because most systems will have belts close to their host stars around 4 au as shown in figure~\ref{fig:abelthist}, and companions that are around 1000 au as shown in figure~\ref{fig:acomphist}. Hence, the octupole strength $\epsilon$ will be extremely weak and only some of these systems will have a large enough mutual inclination to undergo extreme Kozai-Lidov oscillations. Furthermore, the timescale for these systems to undego Kozai-Lidov will be long (equation~\ref{eq:octtimemetzger}) such that, over this period of time, assuming the stellar system has not left the main sequence and ended its life, the close-in belt will have collisionally ground away leaving it with a very low mass.

\begin{figure}
    \centering
    \includegraphics[width=\columnwidth]{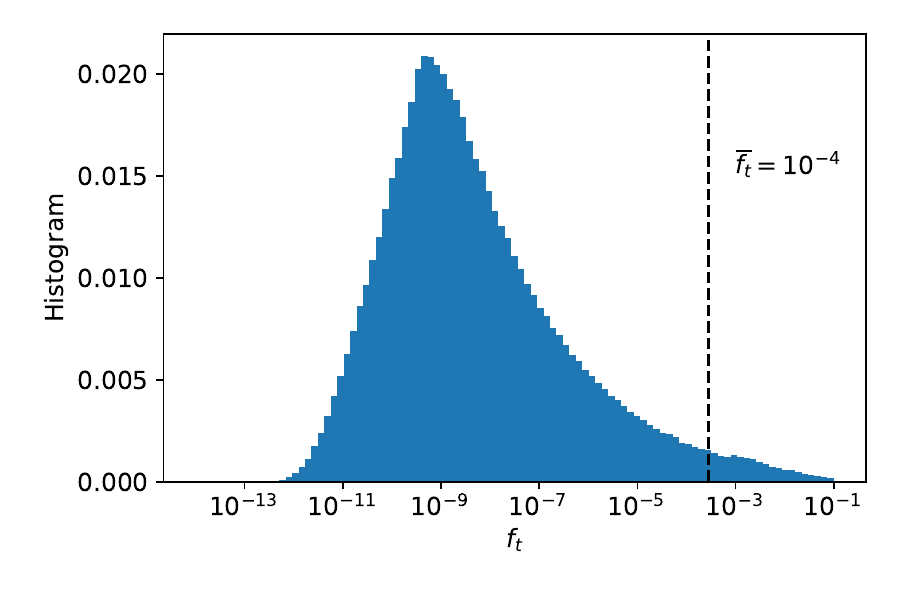}
    \caption{Histogram of all the non-negative values of $f_{\rm{t}}$ of the systems that survived all the cuts. For those that did survive, $8\%$ had non-zero values of $f_{\rm{t}}$ and are shown here. The mean value of $f_{\rm{t}}$ is shown by the dashed black line and is clearly skewed by the highest values such that the vast majority of systems have a value of $f_{\rm{t}}$ that lie below this value.}
    \label{fig:fthist}
\end{figure}

\subsection{Importance of the EKM vs. the SKM}
Figure~\ref{fig:CorrectlyAlignedSystems} shows the relative importance of including the effects of the EKM as opposed to using the simpler case of the SKM as an approximation. It shows the percentage of systems in the model that have an inclination greater than the critical inclination for their system $i_{\rm{crit}}$ above which all planetesimals in the belt are excited to low scaled pericentres for both the EKM and SKM cases. For the simpler SKM case, $i_{\rm{crit}}$ is calculated using equation~\ref{eq:SKMincrange} and is the same for every system in the Monte Carlo model. Conversely, for the EKM, $i_{\rm{crit}}$ is unique to each system and is calculated using the formalism outlined in section~\ref{section:FracExcited}. It shows that there is a difference between the two cases, albeit slight, and that the EKM does increase the number of systems that have high enough inclinations by about $\sim 3\%$. For the critical scaled pericentre considered in the Monte Carlo model, $\sim 14 \%$ of the systems have a misalignment large enough for the EKM to take effect. The overall percentages in each case depend on the critical scaled pericentre $q'_{\rm{crit}} = 1-e_{\rm{crit}}$ that planetesimals are required to reach: the smaller the value of $q_{\rm{crit}}$ that is needed, the less systems that are correctly aligned. For the lowest scaled pericentres, the difference in the percentage of correctly aligned systems between the SKM and EKM cases can traverse an order of magnitude and hence results will differ significantly depending on which case is used in the modelling. For the EKM, the critical inclination above which most planetesimals are excited to high eccentricities depends on $\epsilon$ and thus on $a_{\rm{b}}$, $a_{\rm{c}}$ and $e_{\rm{comp}}$. Therefore, the percentage of the population that have inclinations above $i_{\rm{crit}}$ depends on the distributions of these parameters and hence on the cuts imposed as these can and do change these distributions as shown in figures~\ref{fig:abelthist} and~\ref{fig:acomphist}.

\begin{figure}
    \centering
    \includegraphics[width=\columnwidth]{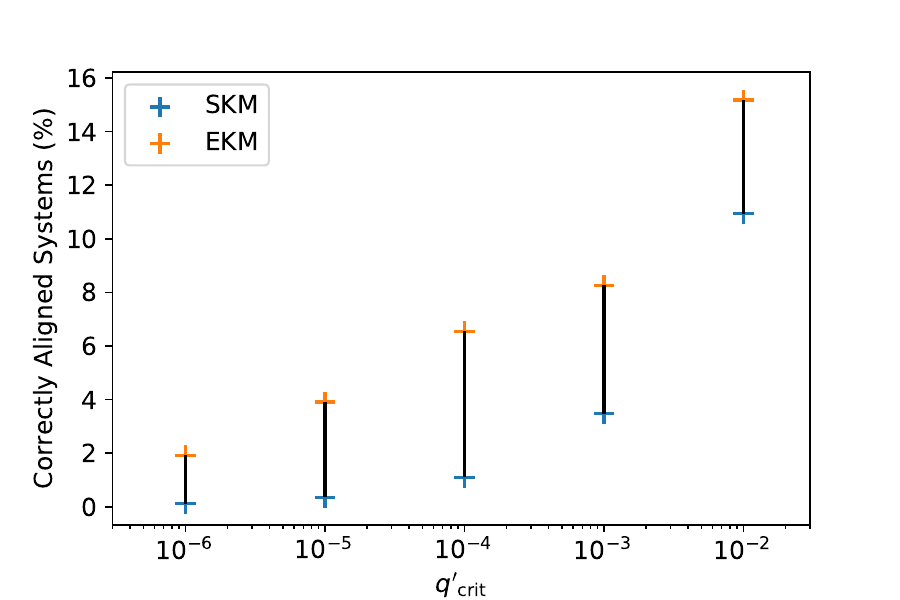}
    \caption{Percentage of systems in the Monte Carlo model whose inclinations exceed those prescribed by the SKM (blue points) and the EKM (orange points) to send particles beyond a threshold eccentricity as a function of that threshold value. Black lines show, for each threshold eccentricity, the difference of the percentage of systems in the model that are sufficiently inclined when using SKM and the EKM.}
    \label{fig:CorrectlyAlignedSystems}
\end{figure}

\section{Discussion}
\label{section:discussion}
The likelihood of the Kozai mechanism as the origin of the observations of KIC 8462852 is small but not entirely improbable. The Monte Carlo simulation shows that, for a Kepler-like distribution of stars, the expected observed rate of stars with planetesimals excited to high eccentricities is $\num{6.6e-9}$. This arises because, from figure~\ref{fig:BeltandCompLoc}, the most likely systems to be seen transiting are those with belts and binary companions where $10^2 \rm{au} \lesssim a_{\rm{b}} \lesssim 10^3 \rm{au}$ and $10^2 \rm{au} \lesssim a_{\rm{c}} \lesssim 10^4 \rm{au}$, which are approximately $1\%$ of systems. Only $\sim 14\%$ of these systems have a large enough inclination for the eccentric Kozai mechanism to take effect and, for those that do, they spend, on average, $0.08\%$ of their main sequence lifetimes in the transient state where large objects are excited to high eccentricities. Not all of these would be observable in the form of dips in their lightcurves, however, as the orbits would need to be correctly aligned with the line of sight from earth and this geometrical transit probability is approximately $0.8\%$. Taken together, this accounts for the calculated expected rate of $\sim 10^{-8}$ that is the output of the model. The model also shows that the most likely belts to undergo this behaviour are like those seen in observations of debris disc systems with $10^2 < a_{\rm{b}}/\rm{au} < 10^3$. Additionally, the companions that are most likely to cause belts to undergo this instability are at intermediate distances for wide binaries: at around 100s-1000s of au. This matches the observed projected separation of the companion star of KIC 8462852, found by \citet{2021ApJ...909..216P} to be 878 $\pm$ 8 au. \newline 

Care should be taken with this, however, as the measurement by \citet{2021ApJ...909..216P} is only the projected on sky separation between KIC 8462852 and the M dwarf and not necessarily the semi-major axis of its orbit. Figure~\ref{fig:ahist} shows the distribution of possible semi-major axes that are consistent with the observed projected separation \citep{YelvertonBinaries}. The distribution was calculated by producing separations calculated from random orbits with uniformly distributed random values of $i$, $\Omega$, $e$ and mean anomaly M. The semi-major axes are derived from the same log normal period distribution that is used in the Monte Carlo model, that was the best fit to observations of wide binaries \citep{2010ApJS..190....1R}. Orbits were considered to have produced a correct separation on a probabilistic basis, with the probability of acceptance depending on the produced separation itself and given by a Gaussian centred on 878 au with a standard deviation of 8 au. Figure~\ref{fig:ahist} shows that the possible semi-major axes of the companion range from 439 to 2000-3000 au. The lower limit arises because orbits with lower $a$ would not reach a separation of 878 au even with $e \approx 1$, whilst the tail is due to orbits with larger $a$ needing more eccentric or edge on orbits to produce the correct separation. Hence, the distribution of possible semi-major axes of the M-dwarf companion is still consistent with the range of semi-major axes of wide binaries most likely to induce the Kozai instability in planetesimal belts. \newline

\begin{figure}
    \centering
    \includegraphics[width=\columnwidth]{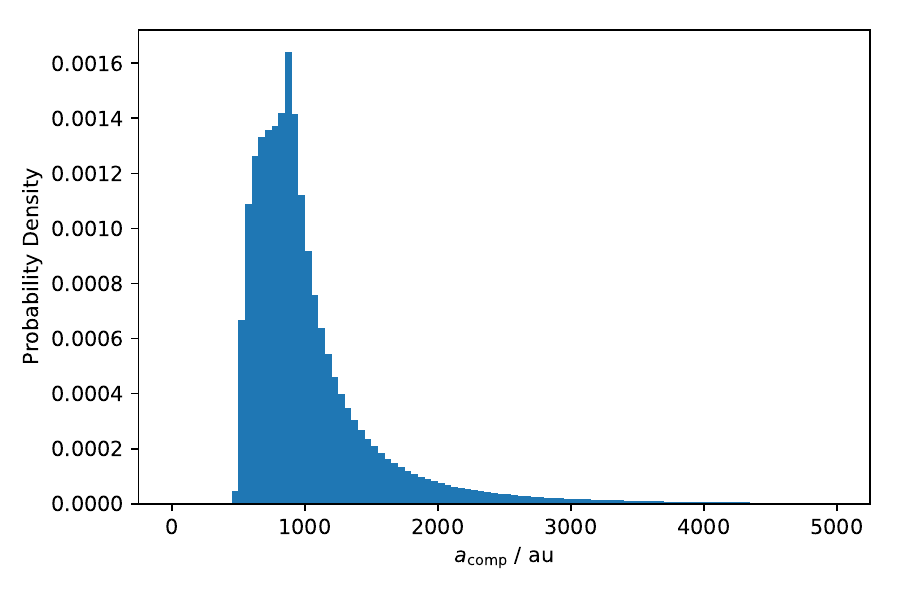}
    \caption{Probability density function of the possible semi-major axes of the M dwarf companion's orbit, given its observed separation of (878 $\pm$ 8) au from KIC 8462852. Assumed priors on the companion's orbit are: randomly distributed $\omega$, randomly distributed $\cos{i}$, randomly distributed $e$, and semi-major axes drawn from the lognormal period distribution of ~\citet{2010ApJS..190....1R}.}
    \label{fig:ahist}
\end{figure}

\subsection{Dependence on Model Parameters and Distributions}

The Monte Carlo model that has been built, and hence the results, depends on a certain number of parameters whose true values are unknown. The most important of these is the `duration of transiting events' $t_{\rm{dur}}$ and the dependence of the number of stars in the Kepler field expected to show KIC 8462852-like dips, $\langle N_{\rm{exp}} \rangle$, on this parameter is shown in figure~\ref{fig:Nstarvstdur}. It is clear that, the longer the transiting events last for, the greater the probability of observing a star with a KIC 8462852-like light-curve. However, they are not proportional to each other as would be expected from equation~\ref{eq:ftunsat} and this is because this equation only holds for those systems that are in the unsaturated state. As $t_{\rm{dur}}$ increases so too does the percentage of saturated systems and, as the value of $\overline{f_{\rm{t}}}$ for saturated systems is independent of $t_{\rm{dur}}$ when $t_{\rm{dur}}$ is small, this increase accounts for the shallower relationship between $\langle N_{\rm{exp}} \rangle$ and $t_{\rm{dur}}$ that would otherwise be expected. \newline

\begin{figure}
    \centering
    \includegraphics[width=\columnwidth]{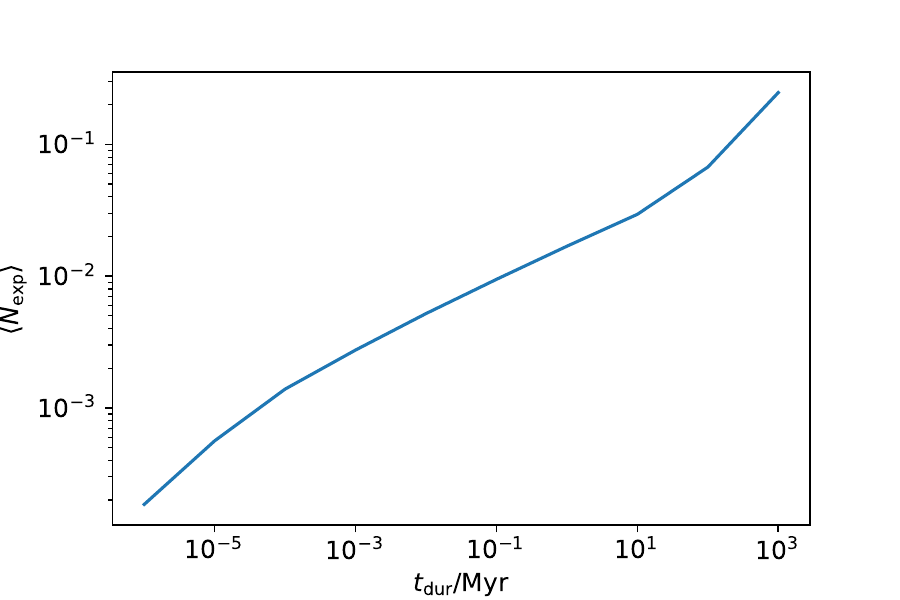}
    \caption{Expected number of KIC 8462852-like objects in the Kepler field as a function of the duration of the observable transit signature caused by each breakup event of a parent body  with a mass greater than $10^{-6} M_{\oplus}$.}
    \label{fig:Nstarvstdur}
\end{figure}

The value of $t_{\rm{dur}}$ doesn't just affect the expected number of KIC 8462852-like stars, it also affects the most likely parameters of observable systems. For example, figures~\ref{fig:ft_funcofabelt_manytdur} and~\ref{fig:ft_funcofacomp_manytdur} show the most likely belt radii and companion semi-major axes to be observed respectively for three different values of $t_{\rm{dur}}$. For small values of $t_{\rm{dur}}$ (i.e. 1-100 yr) only the most distant belts and companions are expected to be observed. However, if $t_{\rm{dur}}$ is increased to an extreme value of 1 Myr, then a large range of belts (10-1000 au) and companions (300-10000 au) are likely to be observed. Similarly, figure~\ref{fig:agevsft} shows how the most likely age of observed systems changes with $t_{\rm{dur}}$; though the age is less sensitive to this free parameter, the smallest values of $t_{\rm{dur}}$ tend to disfavour the oldest systems.
\newline

\begin{figure}
    \centering
    \includegraphics[width=\columnwidth]{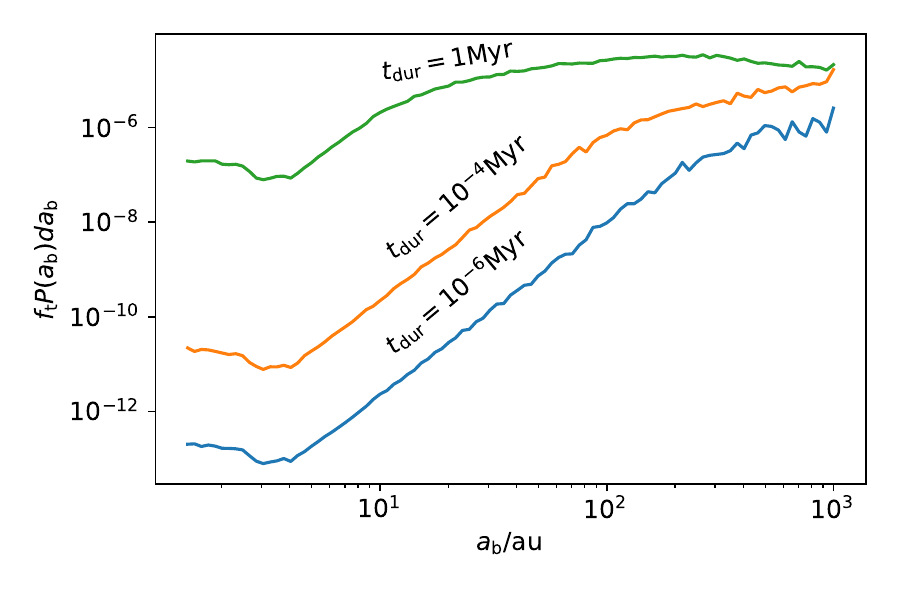}
    \caption{Fraction of stellar lifetime spent with large ($m > 10^{-6} M_{\oplus}$) bodies at small pericentres as a function of belt radius $a_{\rm{b}}$, weighted by the probability that a belt would be found there. The probability distribution of belts used is that of the post-cut population (orange histogram in figure~\ref{fig:abelthist}). The different colour curves represent different values of $t_{\rm{dur}}$, the lifetime of observable transits caused by the breakup of massive bodies on sufficiently eccentric orbits, in Myr.}
    \label{fig:ft_funcofabelt_manytdur}
\end{figure}

\begin{figure}
    \centering
    \includegraphics[width=\columnwidth]{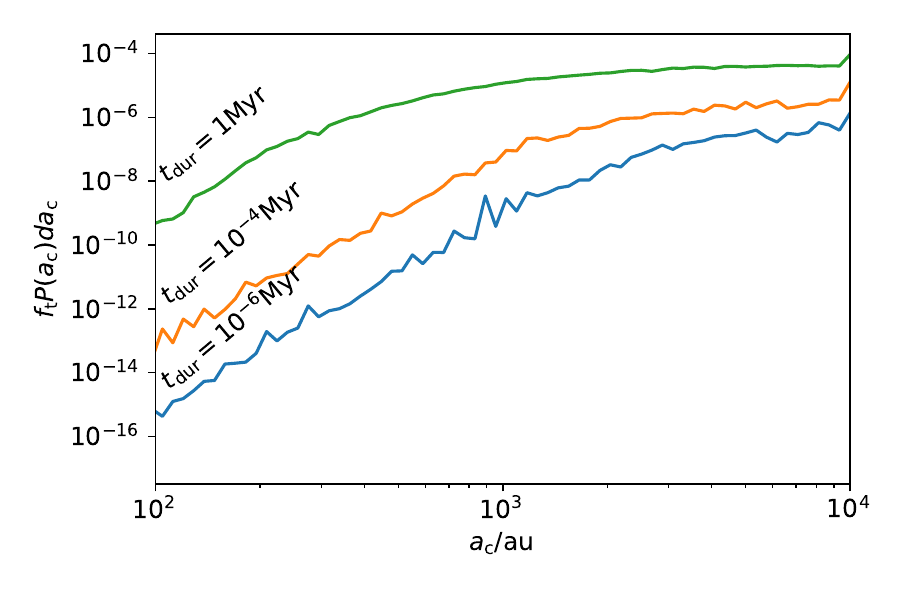}
    \caption{Fraction of stellar lifetime spent with large ($m > 10^{-6} M_{\oplus}$) bodies at small pericentres as a function of companion semi-major axis $a_{\rm{c}}$, weighted by the probability that a companion would be found to have that semi-major axis. The probability distribution used is that of the post-cut population (orange histogram in figure~\ref{fig:acomphist}). The different colour curves represent different values of $t_{\rm{dur}}$, the lifetime of observable transits caused by the breakup of massive bodies on sufficiently eccentric orbits, in Myr.}
    \label{fig:ft_funcofacomp_manytdur}
\end{figure}

The value of $t_{\rm{dur}}$ reflects the lifetime of dust on an eccentric orbit around a central star and hence for how long any optical dips would be observable. The Kreutz family are highly inclined and eccentric sungrazing comets in our own system that are the result of breakups of larger parent bodies, albeit orders of magnitude smaller than the parent body hypothesised for the KIC 8462852 system \citep{KreutzOriginalPaper}. These have been observed for hundreds of years and have orbital periods of $10^2-10^3$ years and hence must have lifetimes of many orbital periods ($\sim 10^3$ yr) \citep{KreutzFamilyLifetime}. Additionally, constraints on the lifetime of large dust releasing bodies can be found using the observations of the depth of optical dips as measured by \citet{2016MNRAS.457.3988B}. \newline

We consider a comet of mass $M_{\rm{comet}}$, density $\rho_{\rm{comet}}$ and radius $R_{\rm{comet}}$ at the pericentre of its orbit at distance $r_{\rm{p}}$ from the central star and which is emitting dust as a spherically symmetric wind. Mass conservation implies that for a constant mass loss rate $\dot{M}$

\begin{equation}
    \label{eq:MassLoss}
    \dot{M} = 4 \pi r^2 \rho_0 \left(\frac{r}{r_0}\right)^{-2} u,
\end{equation}

\noindent where r is radial distance from the comet, $\rho_0$ and $r_0$ are the density and radius respectively at some reference position and $u$ is the speed of the dust. \newline

The depth of the optical dips measured around KIC 8462852 $\delta$ caused by material of optical depth $\tau$ covering a fraction $\Omega_*$ of the stellar surface is 

\begin{equation}
    \label{eq:dipdepth}
    \delta = \tau \Omega_*,
\end{equation}

\noindent for $\tau \ll 1$ and where optical depth is itself given by the line of sight (z axis) absorption due to material with an opacity $\kappa$ i.e.

\begin{equation}
    \label{eq:opticaldepth}
    \tau = \int \kappa \rho dz.
\end{equation}

\noindent The opacity $\kappa$ is the ratio of the interaction cross section of a particle to its mass which, assuming a dust size $s$ and density $\rho_{\rm{d}}$, is 

\begin{equation}
    \label{eq:Kappa}
    \kappa = \frac{3}{4 s \rho_{\rm{d}}}.
\end{equation}

\noindent Using equation~\ref{eq:opticaldepth} and considering the star as a point source, if the comet is transiting with impact parameter $b=0$ and speed $v$ along the $x$ axis such that when $t=0$ then $x=0$, then, at $t=0$ which corresponds to the deepest part of the dip and assuming the size of the clump is approximately $r_{\rm{p}}$

\begin{equation}
    \label{eq:opticaldepthnew}
    \tau = \frac{3 \rho_0 r_0^2}{2s \rho_{\rm{d}} r_{\rm{p}}}.
\end{equation}

\noindent Using equations~\ref{eq:dipdepth} and~\ref{eq:opticaldepthnew}, the reference density and radius can be related to the dip depth by

\begin{equation}
    \label{eq:refdensityandradius}
    \rho_0 r_0^2 = \frac{2s \rho_{\rm{d}} r_{\rm{p}} \delta}{3},
\end{equation}

\noindent where $\Omega_* = 1$ has been used as the star is considered to be a point source in this approximation. Substituting equation~\ref{eq:refdensityandradius} into~\ref{eq:MassLoss} and further assuming that the velocity is approximately the escape velocity of the comet $u_{\rm{esc}} = \sqrt{\frac{8}{3} \pi G \rho_{\rm{comet}}} R_{\rm{comet}}$ gives an expession for the mass loss rate in terms of the dip depth

\begin{equation}
    \dot{M} = \frac{8 \pi s \rho_{\rm{d}} r_{\rm{p}} \delta}{3}  \sqrt{\frac{8}{3} \pi G \rho_{\rm{comet}}} R_{\rm{comet}}.
\end{equation}

\noindent Hence, assuming $\rho_{\rm{d}} = \rho_{\rm{comet}} = \rho$ and using $\delta = 0.2$ as observed by \citet{2016MNRAS.457.3988B}, the evaporation timescale $t_{\rm{evap}} = \frac{M_{\rm{comet}}}{\dot{M}}$ is

\begin{equation}
    t_{\rm{evap}} = 23 \left( \frac{R_{\rm{comet}}}{100 \rm{km}} \right)^2 \left( \frac{\rho}{2700 \rm{kg} \rm{m}^{-3}} \right)^{-1/2} \left( \frac{s}{1 \mu \rm{m}} \right)^{-1} \left( \frac{r_{\rm{p}}}{0.1 \rm{au}} \right)^{-1} \rm{yr}.
\end{equation}

\noindent This estimate is found using the mass loss rate at pericentre using the depth of the deepest dips observed. However, comets on eccentric orbits only experience mass loss for a small portion of their orbits before they move further from the star towards apocentre where the mass loss rate is much lower and consequently it will take a certain number of orbital periods for the comet to fully evaporate. However, the total time the dip from this one body would be observable for is roughly $t_{\rm{evap}}$ and even if there are multiple evaporating bodies close in orbital phase then $t_{\rm{dur}}$ will still be roughly $t_{\rm{evap}}$ or slightly larger.  \newline

Another model parameter that affects the outcome is $M_{\rm{mid}}$ which is the peak of the log-normal distribution of debris disc masses all stars are assumed to be born with that, along with the maximum size of their planetesimals $D_{\rm{c}}$, is constrained by \citet{2018MNRAS.475.3046S}. The results of the model have been based on a value of $10 M_{\oplus}$ which is derived from protoplanetary disk observations \citep{MmidOrigin}. Whilst this parameter sets the maximum mass of belts in the model and should not be set unphysically high, it has no effect on the value of $\langle N_{\rm{exp}} \rangle$. This is because although $M_{\rm{b}} \propto M_{\rm{mid}}$, the number of objects, per unit belt mass, between $m_{\rm{crit}}$ and $m_{\rm{max}}$ ($n_{\rm{c}}'$) is proportional to $M_{\rm{mid}}^{-1}$. Hence the total number of objects in a belt with masses between $m_{\rm{crit}}$ and $m_{\rm{max}}$ ($n_{\rm{c}}' M_{\rm{b}}$) is independent of $M_{\rm{mid}}$. However, $M_{\rm{mid}}$ does have a minimum value in order for the belts to have planetesimals that are large enough to cause dimming events (i.e. $m_{\rm{max}} > 10^{-6} M_{\oplus}$) and this occurs at $M_{\rm{mid}} = 1.27 M_{\oplus}$. \newline

There are different hypothesised eccentricity distributions for wide binaries whose applicability depends on the formation mechanism of the stars themselves. The difficulty in constraining the eccentricity distribution from observations of wide binaries is due to their very long periods (i.e. a semi-major axis of $\sim$ 900 au corresponds to a period of $\sim$ 20,000 years for solar mass stars), which means that a tiny fraction of an orbital arc is covered by the observations leading to many possible orbits with a wide variety of eccentricities that fit the data. For example \citet{2010ApJS..190....1R} found that the eccentricity distribution was consistent with being uniform. However, other studies by \citet{ThermalEccDist} have found that the eccentricity distribution is thermal (i.e. $\propto$ e) or even super thermal for wide binaries. The model was rerun with these different eccentricity distributions but they did not affect the results as the eccentricity only weakly influences the EKM timescale.

\subsection{Applicability to other dusty Stars}

The eccentric Kozai mechanism is a convenient mechanism for exciting objects to high eccentricities and is often claimed as a potential cause of multiple observed phenomena. For example, various stars are observed to have what is termed `Extreme Debris Discs' (EDDs) which are identifiable by very hot dust close to the star (blackbody radii $R_{\rm{bb}} < 1$ au and fractional luminosities $f>0.01$). This dust could not have formed in situ as it would have collisionally depleted over the age of the stars \citep{WyattDiscModel}, of which the lifetimes of some are found to be greater than 100 Myrs \citep{MoorEDDs,WeinbergerGiantImpactsCalc}. One explanation for this phenomenon is that it is the result of giant impacts where, after planetary embryos are formed and the gas disc dissipates, embryos are dynamically excited onto crossing orbits and collide \citep{GiantImpacts,ChambersGIs}. However, simulations show that the era of giant impacts is $\sim 30-200$ Myr \citep{MoonFormingImpact,EraOfGiantImpacts} which is difficult to reconcile with the ages of the oldest EDD systems. On the other hand, it is not trivial to instead assign the longer timescale Kozai mechanism as the cause of this close-in dust. The results of this work show that, whilst the expected ages of most systems would be 100-1000 Myr, the expected rate is not necessarily applicable to EDDs as the input parameters were taken from those stars in the Kepler field. In order to get a meaningful comparison, the model must be rerun accounting for any biases of the searches for EDDs \citep{KennedyEDDs,KennedyExozodi} which is beyond the scope of this paper. \newline

Similar to EDDs, exozodiacal dust is defined to be warm dust within the habitable zone of a system (though the demarcation between the two is ill defined). \citet{KennedyExozodi} find warm $12 \mu \rm{m}$ excesses are detectable towards $1\%$ of stars with a majority of systems identified around young stars ($<120$ Myr) and that they correlate with cold outer belts like in $\eta$ Corvi \citep{EtaCorviObs}. Some exozodi can be explained by dust from collisions in the outer belt migrating inwards through PR drag \citep{JessPRDrag} but others like $\eta$ Corvi require a scattering chain of planets \citep{EtaCorviScattering} to deliver cometary material inwards through many scattering events which then fragment \citep{JessCometFrag}. Though the EKM is a possible cause of delivery, not all systems with warm exozodi are in known stellar binaries although the possibility of misaligned planets in these systems cannot be discounted. \newline

Exocomets have been found through lightcurve analysis around other stars in the Kepler and TESS samples \citep{Exocometsearch} and most of these systems are consistent with being $\sim 100$ Myr old. Additionally, the presence of exocomets can also be inferred from detecting the gas they release using emission line spectroscopy \citep{Rebollido}. It is possible that the EKM is the cause of some of these observations though the results of this model show that, for the case of wide stellar binary perturbers, it is too rare to explain all the systems. Whilst the model struggles to account for the one star with an odd lightcurve, it is interesting to note that the lightcurve of the recently discovered TESS star TIC 43488669 \citep{Tajiri} shows a remarkably similar lightcurve to KIC 8462852 in terms of its complexity. This would increase the known number of KIC 8462852-like stars and could cause worse agreement between this model and the data, though this model was developed for the Kepler field and not for TESS. \newline

The Kozai mechanism is also claimed to be a likely cause of some observations of White Dwarfs (WDs). A not insignificant proportion of White Dwarfs' atmospheres are found to be polluted with metals \citep{WDPollutionStats}, these must have been accreted recently in the history of the star as they have small sinking timescales that would cause them to sink out of the atmosphere and no longer be observable \citep{WDDiffusionTimescaleFontaine,WDDiffusionTimescalePaquette}. This requires recent accretion of planetesimals or disrupted planetary material onto the star which, as White Dwarfs are Gyrs old, suggests that a recent instability could have occurred in the system. As the timescales for the Kozai mechanism can be Gyrs long, it is often claimed that this could contribute to some of the polluted systems seen, though not all of them \citep{WDPollKozai}. Similarly to the pollution, WD 1856b, one of the few planets found transiting a White Dwarf, is thought to have been influenced by the Kozai mechanism \citep{OConnorWD1856,StephanWD1856Kozai}. This is because the planet's current location would mean that, if it had been there on the main sequence, it would have been consumed by the star as it expanded into a red giant \citep{WD1856Engulfment1}. This system is also not just a binary, but part of a higher order system where the Kozai timescale of the distant stars would be long enough to cause the planet to become excited to high eccentricities and migrate inwards where it tidally circularises after the star has evolved to the White Dwarf stage. Whether the Kozai mechanism is a frequent occurence in white dwarf systems is not clear, as figure~\ref{fig:agevsft} shows that, for the smallest values of $t_{\rm{dur}}$, the most common stars to undergo this mechanism are 100-1000 Myr old and there is a sharp downturn at ages greater than 1 Gyr whereas there is no downturn for larger $t_{\rm{dur}}$. In addition, white dwarf systems evolve such that $M_*$, $a_{\rm{b}}$ and $a_{\rm{c}}$ would all change once the main sequence phase has ended which clouds the picture and like the case with the EDDs the exact results of the occurrence rate from this model are not directly applicable. This work only considers the case of stars that undergo Kozai oscillations within the main sequence lifetime of the system and more work will have to be done to examine the population that Kozai after the main sequence, and the biases of White Dwarf observations would have to be accounted for before any comparison could be made.  \newline

This work has sought to quantify the probability that the dips seen in the lightcurve of KIC 8462852 are due to the breakup of an eccentric comet that has undergone Kozai oscillations due to a stellar companion. Whilst the probability found was low, there is a possibility that the Kozai mechanism could still be the cause, albeit not in the form examined in this work. For example, a planet in the system could induce the Kozai instability if it were sufficiently misaligned from any planetesimal belt. Whilst alignment between planets and belts would be expected from formation scenarios, and this is the case in our own solar system, it is not infeasible to have a misalignment. This is evidenced by giant planets which have been found to be significantly inclined to each other such as in $\pi$ Men \citep{JerryPiMen}, as well as the young HD 106906 system where an exterior, eccentric and inclined Jupiter is warping the belt \citep{HD106906ScatLight,HD106906PlanetDetection}. As, for sensible values of $t_{\rm{dur}}$, the model predicts the occurrence of KIC 8462852-like objects to be rare it is worth asking if this disfavours the interpretation of the data as the breakup of an exocomet onto an eccentric orbit. This is not the case, however, as there are other dynamical mechanisms that can place planetesimals onto highly eccentric orbits. The most appealing mechanism would be scattering of material in an outer belt inwards by a planet or chain of planets as is thought to occur in $\eta$ Corvi \citep{EtaCorviScattering}. This would require a chain of planets in the system and for the architecture of the system to be such that the levels of dust supplied by scattering of parent bodies is roughly constant throughout the age of the system otherwise we would be unlikely to observe it. Similarly, another possible mechanism is the resonant destabilisation of a belt. This also requires the presence of a planet such that the locations of its resonance lie in any cold belt of planetesimals in the system such that the dynamics of any bodies in the belt would be chaotic, achieving high eccentricities over the lifetime of the system \citep{KirkwoodGaps,ExoZodiBeltDestab}.

\subsection{Caveats}
\subsubsection{Planets}
The presence of planets in misaligned wide binary systems would act to suppress the Kozai instability induced by the companion. Perturbations from such planets would drive secular (or, for the right period ratios, resonant) oscillations in the orbits of planetesimals. Ample evidence for the influence of planets on smaller bodies comes from our own Solar system in the form of the Asteroid and Kuiper belts, as well as various comet populations \citep{KirkwoodGaps,KuiperBeltResonances}. This influence is also seen in exoplanetary systems, the comets seen in $\beta$ Pic are thought to be scattered inwards from the planetesimal belt by one of the planets in the system \citep{KieferBetaPicExocomets}, whilst the exozodi in the $\eta$ Corvi system is thought to be due to scattering of comets inward from a cold outer belt by a chain of sufficiently massive planets \citep{SebaCometScattering}. There are also eccentric belts, for example Fomalhaut \citep{OldFomalhautImageMacGregor,NewFomalhautImageGaspar}, as well as those that have warps or gaps, which provide evidence that planets can dominate the evolution of planetesimals around them. If this effect is strong enough, usually meaning that the planetesimals are close enough to the planet(s), then the planetary interaction will have a greater effect than that of the binary companion and this would act to shut off the Kozai mechanism in a manner analogous to General Relativity \citep{InnanenPlanetsprecess}. The planet, however, could itself be affected by the star and increase its eccentricity and the effect of this on the planetesimals orbits is unknown though the evolution of planets under the Kozai mechanism may be subject to tidal considerations which severely complicate the picture. In addition to this, a system of multiple planets with or without a belt can precess as a rigid disc in the presence of a highly misaligned companion star instead of undergoing the eccentric Kozai mechanism and avoid destruction \cite{InnanenPlanetsprecess}.

\subsubsection{Input Distributions}
Throughout this work it has been assumed that the inclination distribution of wide binary companions to planetary systems is uniformly distributed. Recent analysis of astrometric observations by \citet{ChristianInclinationDist} and \citet{BehmardInclinationDist}, however, have revealed the possibility that wide binary companions are biased towards low mutual inclinations. This could be caused by the natural inclination distribution that arises out of binary star formation through core fragmentation. Though some binaries would inevitably be formed by capture and have random orientations, these may be in the minority of total wide binary systems and would be represented only at the widest separations. The observed bias could also, however, be due to the Kozai mechanism itself. If the distribution inherited since birth is uniform, then it could be expected that some systems will have a high enough inclination that they will become unstable due to the EKM and hence will not be included in the samples analysed by \citet{ChristianInclinationDist} and \citet{BehmardInclinationDist}, as they will have been destroyed. Though it should be noted that, even for the most highly inclined systems, the susceptibility to the EKM is subject to the same restrictions outlined in section~\ref{section:cuts}. \newline

The parameter distributions used in this model are uncorrelated which is not necessarily true in real systems. For example, more massive stars might be expected to form with more massive protoplanetary discs and hence have more massive debris discs. Similarly, more distant binary companions are more likely to have formed by capture than core fragmentation than close in pairs and thus could be expected to have larger eccentricities. Whilst these would not change the final answer by orders of magnitude, they might affect the most likely masses and ages of stars that would be seen to be undergoing these events.

\section{Conclusions}
\label{section:conclusions}
This work has sought to examine the effect of highly misaligned wide binary companion stars on planetesimal belts, with a specific focus on explaining the extreme lightcurve of KIC 8462852 through the `Eccentric Kozai Mechanism'. The secular equations of motion for the hierarchical three body problem were integrated to show that planetesimals in a belt can reach eccentricities greater than 0.99 for large enough inclinations. The exact inclination above which this occurs depends on the semi-major axes of the planetesimal and companion, but in some cases can be as low as $45^{\circ}$. For these inclinations, not only does this high eccentricity / low pericentre space become unlocked but the integrations also show that, on average, $100\%$ of the belt particles will reach these high eccentricities. \newline

These results were then fed into a Monte Carlo model of the Kepler field that sought to constrain how often the eccentric Kozai mechanism would be expected to produce an observable exocomet signature in the lightcurves of stars and the parameters of the most likely systems to be seen in this state. It was found that the binary systems most likely to be observed with large objects at small pericentres are those with belts at $10^2-10^3$ au, companions at $10^2-10^4$ au, host stars with masses $M_* \geq 1 M_{\odot}$ and stellar ages of $10^2-10^3$ Myr and, apart from the non-detection of a distant belt, all of these parameters match with what is known about the KIC 8462852 system. However the model found, on average, the fraction of their main sequence lifetimes that stars spend with large objects excited to high eccentricities is $\num{2.7e-4}$, with a spread between $10^{-9}-10^{-1}$. This leads to a probability of observing one or more  Kepler stars to have KIC 8462852-like dimming events due to this mechanism of $\num{1.3e-3}$. Hence, though it is possible that the Kozai mechanism might be the cause, it is much more likely than not that another mechanism is responsible, such as scattering by one or more planets undergoing a dynamical instability or resonant destabilisation of planetesimals in a belt. This has potential consequences beyond the interpretation of KIC 8462852 as the eccentric Kozai mechanism is often invoked to explain phenomena such as extreme debris discs. Only by extending this model to these other scenarios can it be determined whether this mechanism occurs often enough to be a viable explanation.  \newline

\section*{Acknowledgements}

SDY thanks the Science and Technology Facilities Council (STFC) for a PhD studentship.

%%%%%%%%%%%%%%%%%%%%%%%%%%%%%%%%%%%%%%%%%%%%%%%%%%
\section*{Data Availability}

This work makes use of the mass distribution of stars in the Kepler data which can be found at \url{https://exoplanetarchive.ipac.caltech.edu/docs/KeplerMission.html}. Additionally, the N-body simulations were carried out using \textsc{rebound} which is freely available at \url{https://rebound.readthedocs.io/en/latest/}.



\bibliographystyle{mnras}
\bibliography{example} % if your bibtex file is called example.bib

%\begin{thebibliography}{99}
%\end{thebibliography}

%%%%%%%%%%%%%%%%%%%%%%%%%%%%%%%%%%%%%%%%%%%%%%%%%%

%%%%%%%%%%%%%%%%% APPENDICES %%%%%%%%%%%%%%%%%%%%%

\appendix

\section{The Secular Equations of Motion}
\label{section:KozaiEqs}

\subsection{The Quadrupole and Octupole Terms in the Disturbing Function}
\label{section:QuadOctTerms}

The quadrupole and octupole terms in the disturbing function are

\begin{equation}
\label{eq:Fquad}
    F_{\rm{quad}} = -\frac{e^2}{2} + \theta^2 + \frac{3}{2}e^2 \theta^2 + \frac{5}{2}e^2(1 - \theta^2)\cos(2\omega),
\end{equation}

\noindent and

\begin{equation}
\label{eq:Foct}
\begin{split}
    F_{\rm{oct}} = \frac{5}{16}(e + \frac{3}{4}e^2)[(1-11\theta-5\theta^2+15\theta^3)\cos(\omega-\Omega) + \\ (1+11\theta-5\theta^2-15\theta^3)\cos(\omega+\Omega)]  - \\ \frac{175}{64}e^3[(1-\theta-\theta^2+\theta^3)\cos(3\omega - \Omega) + \\ (1+\theta-\theta^2-\theta^3)\cos(3\omega+\Omega)],
\end{split}
\end{equation}
\noindent where $\theta = \cos(i)$.

\subsection{The Standard Kozai-Lidov Mechanism}
The time derivatives of the orbital elements of $m_1$ in the SKM case are

\begin{equation}
    \frac{di}{d\tau} = -\frac{15}{8} \frac{e^2}{\sqrt{1-e^2}} \sin(2\omega)\sin(i)\cos(i),
    \label{eq:SKMI}
\end{equation}

\begin{equation}
    \frac{de}{d\tau} = \frac{15}{8} e\sqrt{1-e^2} \sin(2\omega)\sin(2i),
    \label{eq:SKMe}
\end{equation}

\begin{equation}
    \frac{d\omega}{d\tau} = \frac{3}{4} \frac{1}{\sqrt{1-e^2}}[2(1-e^2) + 5\sin^2(\omega)(e^2 - \sin^2i)],
    \label{eq:SKMom}
\end{equation}

\begin{equation}
    \frac{d\Omega}{d\tau} = -\frac{\cos(i)}{4\sqrt{(1-e^2)}}(3+12e^2-15e^2\cos^2(\omega)).
    \label{eq:SKMOm}
\end{equation}

\subsection{The Eccentric Kozai-Lidov Mechanism}
The time derivatives of the orbital elements of $m_1$ in the EKM case are

\begin{equation}
    \frac{di}{d\tau} = \frac{-1}{\sin(i) \sqrt{1-e^2}} \left( \frac{\partial F}{\partial \Omega} - \theta \frac{\partial F}{\partial \omega}\right),
\end{equation}

\begin{equation}
    \frac{de}{d\tau} = \frac{-(1-e^2)^{1/2}}{e} \frac{\partial F}{\partial \omega},
\end{equation}

\begin{equation}
    \frac{d\omega}{d\tau} = \frac{(1-e^2)^{1/2}}{e} \frac{\partial F}{\partial e} + \theta (1-e^2)^{-1/2} \: \frac{\partial F}{\partial \theta},
\end{equation}

\begin{equation}
    \frac{d \Omega}{d \tau} = -(1-e^2)^{-1/2} \: \frac{\partial F}{\partial \theta},
\end{equation}

\noindent where $F = F_{\rm{quad}} + \epsilon F_{\rm{oct}}$.

\newpage
\section{Table of Parameters}

\begin{table*}
    \centering
    \caption{The parameters used in the paper.}
    \begin{tabular}{llc}
         \hline
         Parameter & Meaning & Definition Equation Number    \\
         \hline
         $H^{\rm{TP}}$ & Hamiltonian of a test particle in the prescence of an external, misaligned perturber, expanded to octupole order  & \ref{eq:Hamiltonian} \\
         $F_{\rm{quad}}$ & Quadrupolar term of the Hamiltonian & \ref{eq:Fquad} \\
         $F_{\rm{oct}}$ & Octupolar term of the Hamiltonian & \ref{eq:Foct} \\
         $\epsilon$ & Strength of the octupole terms relative to the quadrupole terms & \ref{eq:epsilon} \\
         $M_*$ & Mass of central star that is orbited by a planetesimal and companion star \\
         $M_{\rm{c}}$ & Mass of the companion star \\
         $a_{\rm{c}}$ & Semi-major axis of the companion star's orbit \\
         $e_{\rm{c}}$ & Eccentricity of the companion star's orbit \\
         $a_{\rm{pl}}$ & Semi-major axis of a massless planetesimal \\
         $e_{\rm{pl}}$ & Eccentricity of a massless planetesimal \\
         $i_{\rm{pl}}$ & Inclination of a massless planetesimal \\
         $\omega_{\rm{pl}}$ & Longitude of pericentre of a massless planetesimal \\
         $\Omega_{\rm{pl}}$ & Longitude of ascending node of a massless planetesimal \\
         $\Omega_*$ & Angular velocity of a massless planetesimal about its host star \\
         $a_{\rm{b}}$ & Semi-major axis of the midpoint of a planetesimal belt \\
         $a_{\rm{b,lower}}$ & Semi-major axis of the inner edge of a planetesimal belt \\
         $a_{\rm{b,upper}}$ & Semi-major axis of the outer edge of a planetesimal belt \\
         $t_{\rm{quad}}$ & Timescale for quadrupolar oscillations & \ref{eq:quadtime}\\
         $t_{\rm{oct}}$ & Timescale for octupolar oscillations & \ref{eq:octtimemetzger} \\
         $q'$ & Scaled pericentre of an orbit, true pericentre divided by semi-major axis & \ref{eq:scaledperidef} \\
         $q'_{\rm{crit}}$ & Critical scaled pericentre below which planetesimals break up and create a KIC 8462852-like light curve \\
         $i_{\rm{crit}}$ & Critical inclination above which planetesimals break up and create a KIC 8462852-like light curve \\
         $F(q'<q'_{\rm{crit}})$ & Fraction of particles in a belt that reach a scaled pericentre less than $q'_{\rm{crit}}$ & \ref{eq:critscaledperi} \\
         $i_{\rm{mid}}$ & Inclination at which $F(q'<q'_{\rm{crit}}) = 1/2$ \\
         $\langle N_{\rm{exp}} \rangle$ & Expected number of observable KIC 8462852-like objects in the Kepler field & \ref{eq:Ntab} \\
         $f_{\rm{reject}}$ & Fraction of the initial sample of the MC model that will not Kozai for physical reasons \\
         $f_{\rm{t}}$ & Fraction of a star's lifetime that it produces an observable, KIC 8462852-like light curve   & \ref{eq:Saturatedft} \& \ref{eq:Unsaturatedft}\\ 
         $P_{\rm{geo}}$ & Geometric transit factor accounting for percentage of orbits crossing the line of sight & \ref{eq:Pgeo} \\
         $p$ & The probability that a star would be observed to have a KIC 8462852-like light curve & \ref{eq:p}\\
         $t_{\rm{MS}}$ & The main sequence lifetime of a star & \ref{eq:MSlifetime} \\
         $N(m>m_{\rm{crit}})$ & The number of planetesimals with a mass greater than $m_{\rm{crit}}$ & \ref{eq:ngreatermcrit} \\
         $t_{\rm{dur}}$ & The length of time a KIC 8462853-like light curve lasts for after the breakup of a large planetesimal \\ 
         $n(D)$ & The size distribution of particles in a collisional cascade & \ref{eq:SizeDist}\\
         $R_{\rm{bb}}$ & The radius of a planetesimal belt if it emitted as a black body & \ref{eq:PRbb}\\
         $M_{\rm{bb}}$ & The mass of a planetesimal belt if its semi-major axis is its black body radius & \ref{eq:Mblackbody} \\
         $M_{\rm{mid}}$ & The peak of the log normal distribution of debris disc masses \\
         $\Gamma$ & The ratio between the black body radius of a disc and its true radius & \ref{eq:gamma} \\
         $M_{\rm{b}}$ & The mass of a planetesimal belt & \ref{eq:Mbelt} \\
         $m_{\rm{max}}$ & The mass of the largest planetesimal in a belt \\
         $\Delta a_{\rm{b}}$ & The width of a planetesimal belt \\
         $t_{\rm{orb,comp}}$ & The orbital period of the companion star \\

         \hline
    \end{tabular}
    \label{Tab:table}
\end{table*}

%%%%%%%%%%%%%%%%%%%%%%%%%%%%%%%%%%%%%%%%%%%%%%%%%%

% Don't change these lines
\bsp	% typesetting comment
\label{lastpage}
\end{document}